\documentclass{article}

\PassOptionsToPackage{numbers}{natbib}
\usepackage[preprint]{neurips_data_2023}

\usepackage[utf8]{inputenc} 
\usepackage[T1]{fontenc}    
\usepackage{hyperref}       
\usepackage{url}            
\usepackage{booktabs}       
\usepackage{amsfonts}       
\usepackage{nicefrac}       
\usepackage{microtype}      
\usepackage{xcolor}         
\usepackage{graphicx}
\usepackage{subcaption}
\usepackage{footmisc}
\usepackage{multirow}
\usepackage{pdfpages}
\usepackage{amsmath} 
\usepackage{verbatim}

\title{Unraveling Molecular Structure:\\ A Multimodal Spectroscopic Dataset for Chemistry}

\author{%
Marvin Alberts$^{1,2,4*}$ \quad Oliver Schilter$^{1,3,4*}$ \quad Federico Zipoli$^{1,4}$ \\
\textbf{Nina Hartrampf}$^2$ \quad \textbf{Teodoro Laino}$^{1,4}$ \\
$^1$IBM Research \quad $^2$University of Zürich \quad $^3$EPFL \quad $^4$NCCR Catalysis\\
\texttt{marvin.alberts@ibm.com}\\
\texttt{\{oli, fzi, teo\}@zurich.ibm.com}\\
\texttt{nina.hartrampf@chem.uzh.ch}\\
}

\begin{document}
\def\thefootnote{$^{\dagger}$}\footnotetext{Equal Contributions}\def\thefootnote{\arabic{footnote}}
\maketitle

\begin{abstract}
Spectroscopic techniques are essential tools for determining the structure of molecules. Different spectroscopic techniques, such as Nuclear magnetic resonance (NMR), Infrared spectroscopy, and Mass Spectrometry, provide insight into the molecular structure, including the presence or absence of functional groups. Chemists leverage the complementary nature of the different methods to their advantage. However, the lack of a comprehensive multimodal dataset, containing spectra from a variety of spectroscopic techniques, has limited machine-learning approaches mostly to single-modality tasks for predicting molecular structures from spectra. 
Here we introduce a dataset comprising simulated $^1$H-NMR, $^{13}$C-NMR, HSQC-NMR, Infrared, and Mass spectra (positive and negative ion modes) for 790k molecules extracted from chemical reactions in patent data. This dataset enables the development of foundation models for integrating information from multiple spectroscopic modalities, emulating the approach employed by human experts. Additionally, we provide benchmarks for evaluating single-modality tasks such as structure elucidation, predicting the spectra for a target molecule, and functional group predictions. 
This dataset has the potential automate structure elucidation, streamlining the molecular discovery pipeline from synthesis to structure determination. 
The dataset and code for the benchmarks can be found at \url{https://rxn4chemistry.github.io/multimodal-spectroscopic-dataset}.
\end{abstract}

\section{Introduction}
The rapid advancement of artificial intelligence (AI) and machine learning (ML) methods has ushered in a new era for the field of chemistry. Computational approaches have transformed various aspects of chemical research, including retrosynthesis planning \cite{segler2018planning,coley2018machine,schwaller2020predicting,genheden2020aizynthfinder}, reaction optimization through Bayesian optimization \cite{mockus1984bayesian,taylor2023accelerated,guo2023bayesian,pyzer2022accelerating}, molecular design \cite{G_mez_Bombarelli_2018,jin2018junction,born2021paccmannrl,manica2023accelerating} and more. Tasks that were previously laborious and time-consuming when performed manually are now being automated, accelerating the discovery process.
Despite these advancements, one critical aspect of chemistry that remains heavily reliant on human expertise is structural elucidation – the process of determining the molecular structure from spectroscopic data. 

While chemists often have an intuition about a molecule that was synthesized, the actual composition of the product needs to be verified using spectroscopic data. Different spectroscopic techniques yield different types of information. For instance, certain functional groups (e.g., alcohols) will exhibit characteristic peaks in specific regions of the infrared (IR) spectrum (e.g., 3200-3300 cm$^{-1}$ \cite{workman2007practical}), while the mass spectrum (MS) can be used to find the molecular weight of a molecule in question. Similar to solving a complex puzzle, the more spectroscopic modalities a chemist has access to, the more information and hints they can gather to predict the molecular structure and explain the observed spectral peaks. 

While AI/ML models have been developed for this task, they predominantly focus on single spectroscopic modalities, such as infrared (IR) \cite{alberts2023leveraging,zhang2022review} or nuclear magnetic resonance (NMR) \cite{huang2021framework,alberts2023learning} spectroscopy. In contrast, human experts leverage multiple modalities by combining information from various spectroscopic techniques to gain a better understanding of the molecular structure.
To bridge this gap and enable the automation of structural elucidation, there is a need for a multimodal dataset containing spectra from a variety of spectroscopic techniques. 

Multimodal datasets in other fields, such as computer vision and natural language processing 
 \cite{changpinyo2021conceptual,eftekhar2021omnidata,roberts2021hypersim,schuhmann2022laion,kakaobrain2022coyo}, have enabled remarkable achievements like text-to-image generation \cite{ramesh2021zero,reed2016generative,rombach2022high}, image captioning \cite{vinyals2015show}, object detection using bounding boxes \cite{redmon2016you,lin2017feature}, and even multitask models \cite{mizrahi20244m,radford2021learning}. Similarly, we postulate that a multimodal dataset for chemical spectra could lead to significant advancements. Such a dataset would serve as a valuable resource for developing AI/ML models capable of integrating information from multiple spectroscopic modalities, emulating the approach employed by human experts in analyzing and interpreting spectral data.

In this paper, we introduce a dataset comprising simulated IR, $^1$H-NMR, $^{13}$C-NMR, Heteronuclear Single Quantum Coherence (HSQC)-NMR, positive-ion mass spectrometry (MS), and negative-ion MS spectra for a large set of 790k realistic molecules extracted from patent data. We specifically sample molecules from the United States Patent Office (USPTO) dataset, which is commonly used for reaction prediction \cite{coley2017prediction,schwaller2019molecular}.
We also introduce initial baseline models for single-modality tasks, namely predicting molecular structures from spectral data, generating spectras from molecular structures, and identifying functional groups present in molecules based on spectral information. These models demonstrate the potential of our dataset for automated molecular structure elucidation and serve as benchmarks for evaluating other AI architectures on these tasks.

By leveraging AI/ML methodologies and the comprehensive information from multiple spectroscopic modalities, this dataset has the potential to close the loop between automated synthesis and automated structural elucidation, streamlining the molecular discovery cycle.
\section{Related Work}

\textbf{USPTO Dataset:}
The USPTO dataset, by \citet{lowe2012extraction}, has become a staple for machine learning based works in chemistry \cite{coley2018machine,coley2017prediction,schwaller2019molecular,toniato2021unassisted}. It is sourced from patent data and in contrast to many other datasets in chemistry fully open source, making a popular choice  for training and evaluating reaction prediction models. The main advantage of this dataset is that all molecules are sourced from patent data, i.e. their distribution is very similar to molecules common in industry and to a lesser extent academia.

\textbf{NMR:} 
Predicting the chemical structure from NMR spectra remains a largely unexplored subject. \citet{jonas2019deep} first utilized imitation learning to predict the molecular structure from $^{13}$C-NMR spectra. \citet{sridharan2022deep} approached the problem from a different angle using a reinforcement learning guided Monte Carlo tree search to generate molecules from $^{13}$C-NMR spectra. The first work to combine both $^1$H- and $^{13}$C-NMR spectra employed 1D-CNNs to predict substructures contained in the parent molecule from both $^1$H and $^{13}$C-NMR spectra. Subsequently, a database search is employed to provide the closest match \cite{huang2021framework}. More recently \citet{alberts2023leveraging} demonstrated that Transformer models are capable of generating molecular structure from annotated NMR spectra. 

However, very few other works have been published and comparison between the approaches is rare. A few studies have evaluated model performance on the experimental spectra available in the nmrshiftdb2 database \cite{kuhn2024twenty} but most works utilise different private datasets. While some exclusively train on a limited amount of experimental spectra, most simulate a large number of spectra and pretrain on these simulated spectra. At the time of writing, none of the simulated datasets used in these works are publicly available hindering the transparent benchmarking of model architectures.

\textbf{IR:} 
Similar to NMR spectra few works investigate full structure elucidation from IR spectra. \citet{alberts2023leveraging} showed that it is possible to predict the chemical structure from IR spectra of small molecules. On the other hand, predicting the presence of certain functional groups from IR spectra has been assessed extensively \cite{fine2020spectral, enders2021functional, wang2023infrared, jung2023automatic}. For this application there are also no standardized datasets with each work using different spectra for training and evaluation. The closest standardised dataset, the NIST Gas-Phase IR Database \cite{stein2008nist} contains solely 5,228 spectra limiting the applicability for machine models. 

\textbf{MS/MS:}
Out of the three spectroscopic methods, structure elucidation from MS/MS spectra is the most explored and commonly used in laboratories. However, most approaches rely on matching a given MS/MS spectra to a large database inherently limited by the size and diversity of the database \cite{beck2024recent}. In another approach a fragmentation tree is derived from the MS/MS spectrum and matched to a database of fragmentation trees. While also relying on a database, this approach is less limited as fragmentation trees can be predicted with relatively high fidelity \cite{bocker2016fragmentation, vaniya2015using, goldman_generating_2024, goldman_prefix-tree_2023}. To remove the need for database matching some works have proposed predicting the chemical structure directly from the MS/MS spectrum. Of these MSNovelist \cite{stravs2022msnovelist} relies on an LSTM whereas MassGenie \cite{shrivastava2021massgenie} utilizes a Transformer model to predict the structure as Simplified molecular-input line-entry system (SMILES) \cite{weininger1988smiles}.
As with NMR and IR spectra, the datasets used for training mass spectrometry models are often not publicly accessible. These datasets typically require a commercial license for access, such as the NIST MS database\cite{nist}. However, efforts to create open-access repositories of experimentally measured data are underway, with the GNPS database\cite{wang2016sharing} being a notable example.

\section{Dataset}
\begin{figure}
    \centering
    \includegraphics[width=1\textwidth]{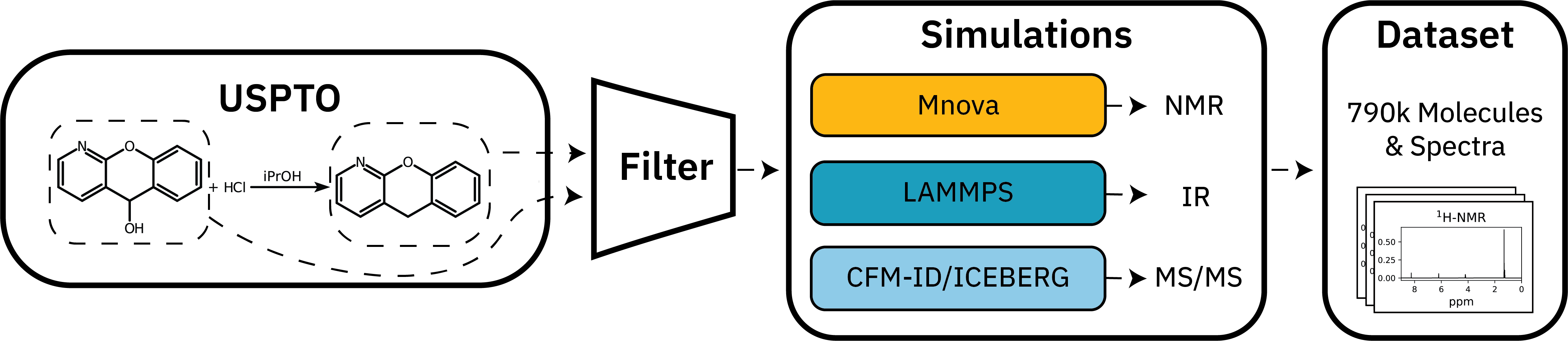}
    \caption{Overall workflow: Molecules are extracted from reaction data (USPTO), filtered to only contain certain atom types as well as minimum and maximum molecule size, then for each molecule the corresponding spectra are simulated resulting in a dataset of spectra for 790k molecules.}
    \label{fig:overview}
\end{figure}
Since in organic chemistry spectral data is often acquired during reactions to monitor progress or after completion, a dataset intended for inferring molecular structures should encompass a chemical space similar to that accessible through common organic chemistry reactions. Therefore, we chose to utilize the USPTO reaction dataset, mined by \citet{lowe2012extraction} who extracted chemical reactions from the US patent database. This dataset spans 1,435,481 chemical reactions across various reaction classes and as such only contains realistic molecular structures and commonly used chemicals such as solvents, reactant and reagent.
We identified all unique molecules from these reactions and applied filtering criteria based on the heavy atom count (all atoms except Hydrogen), retaining only those molecules with more than five and fewer than 35 heavy atoms. Additionally, we filtered out molecules containing elements other than Carbon, Hydrogen, Oxygen, Nitrogen, Sulfur, Phosphorus, Silicon, Boron, and the halogens. This reduced the number of molecules from 1,675,439 to 1,416,499. We attempted to simulate all molecules; however, since some simulations failed for certain molecules, we opted to include only those molecules for which all spectra simulations were successful (see Figure\ref{fig:overview}).

\begin{figure}
    \centering
    \includegraphics[width=0.85\textwidth]{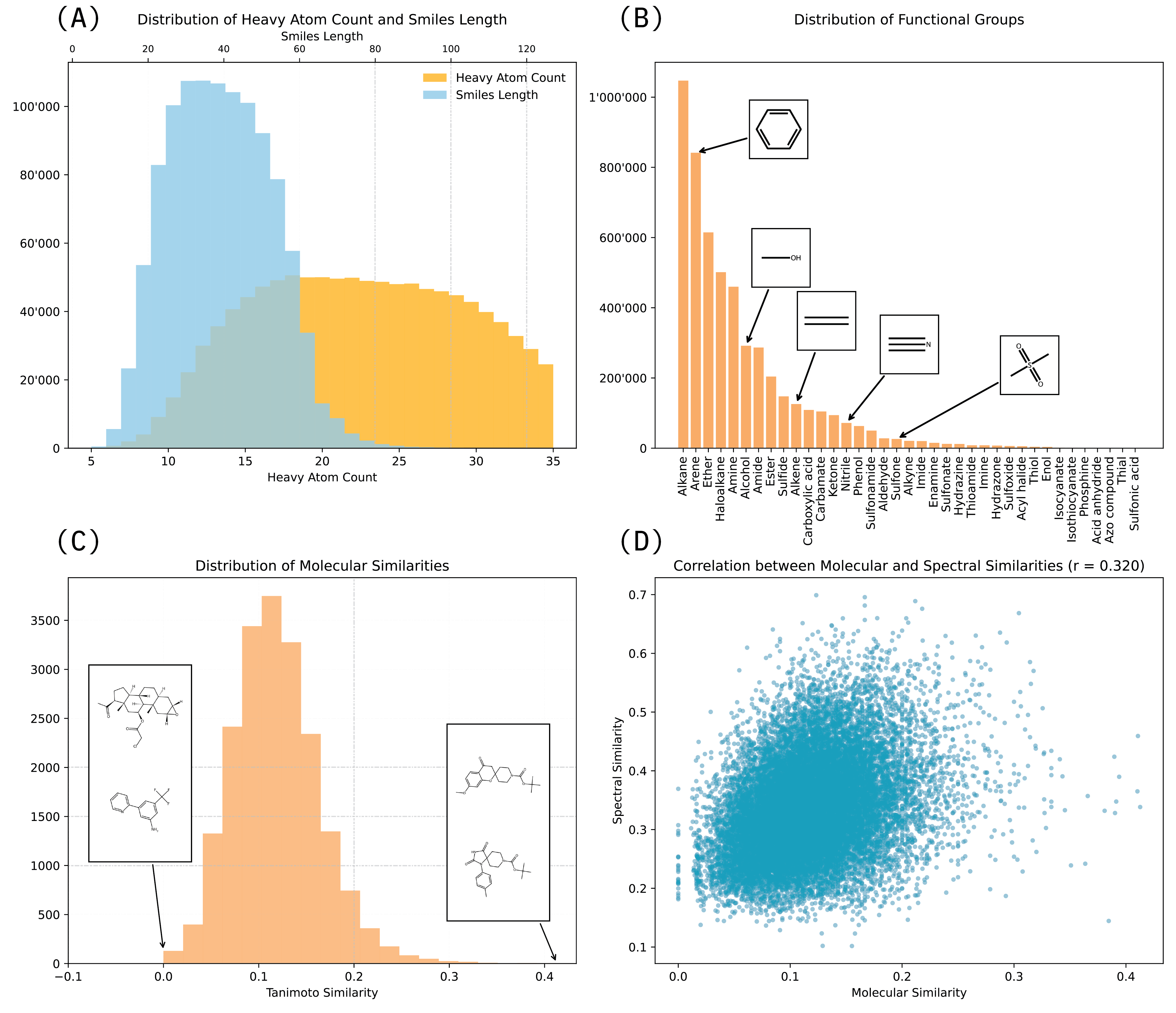}
    \caption{(A) Size  and functional group (B) distribution of the full dataset. 200 randomly sampled molecules were investigated for their chemical similarity to each (C) as well as if the IR spectra similarity correlates to the chemical similarity (D). }
    \label{fig:functional_grousp}
\end{figure}

Overall, we ended up with 794.403 unique molecules and their corresponding IR, $^1$H-NMR, $^{13}$C-NMR, HSQC-NMR, and MS/MS spectra (for more details about the simulations, refer to Section \ref{sec:simulations}). The molecular structures is represented as SMILES and additionally the molecular formula of each molecule (e.g. $C_6H_{12}O_6$) is provided.
The distribution of SMILES lengths and heavy atom counts is visualized in Figure \ref{fig:functional_grousp} (A), spanning the full range between 5 and 35 heavy atoms. Additionally, the chemical similarity between 200 randomly sampled molecules was investigated by calculating the Tanimoto similarity of their chemical fingerprints (see Figure \ref{fig:functional_grousp} (C)). It can be seen that the dataset comprises a broad range of dissimilar chemical structures which is desired.

The chemical similarity is weakly correlated with the similarity in the IR spectra domain, as shown in Figure \ref{fig:functional_grousp} (D), indicating that molecules with similar chemical compositions may also have similar IR spectra. For all similarity calculation refer to section \ref{sec:A_similiarity}. 

As chemical functional groups are often distinctly responsible for patterns in certain areas of the spectra (e.g., aromatic rings causing peaks in the range of 6.0 – 8.7 ppm in the $^1$H-NMR, as exemplified in Section \ref{sec:A_aromatic}), we analyzed the functional group compositions of our collected dataset. In Figure \ref{fig:functional_grousp} (B), the distribution is visualized. As can be seen, the most prevalent functional groups are Alkanes, Arenes, and Ethers, followed by Haloalkanes, and overall spanning a broad range of functional groups.

Overall our dataset comprises 790k unique molecules and their spectra, spanning a large diverse space regarding their chemical similarity, molecule size, as well as functional group composition.

\subsection{Data Generation}
\label{sec:simulations}
An overview of the generated data can be found in Table \ref{tab:datatypes}. In addition, the spectra and annotations for two molecules are shown in Appenxix section \ref{app:sample_spectra}.

\begin{table*}[t]
    \renewcommand{\arraystretch}{1.3}
    \caption{
        Overview of the data available for the different modalities. For all modalities except IR we provide annotations in addition to the unprocessed spectrum.}
    \label{tab:datatypes}
    \begin{center}
    \begin{tabular}{l|c|c}
    \toprule
        Modality & Subtype & Data Description \\
\hline
        IR & Spectrum  & Vector of size 1.800 \\ \hline
        \multirow{2}{0pt}{$^1$H-NMR} &  Spectrum & Vector of size 10.000 \\
        & Annotated Spectrum &  Start, End, Centroid, Integration and Type of each peak  \\\hline
        
        \multirow{2}{0pt}{$^{13}$C-NMR} & Spectrum & Vector of size 10.000 \\
        & Annotated Spectrum &  Centroid and Intensity of each peak\\ \hline
        \multirow{2}{0pt}{HSQC-NMR} & Spectrum & Matrix: 512x512 \\
        & Annotated Spectrum &  X, Y coordinates and integration of each peak\\ \hline
        
        \multirow{2}{0pt}{Positive MS/MS} & Spectrum & m/z \&  Intensity of each peak \\
        & m/z Annotations &  Chemical formula corresponding to the m/z of each peak\\ \hline
        
        \multirow{2}{0pt}{Negative MS/MS} & Spectrum & m/z \& Intensity of each peak \\
        & m/z Annotations &  Chemical formula corresponding to the m/z of each peak\\
    \bottomrule
        \end{tabular}
    \end{center}
\end{table*}

\textbf{NMR Simulations:} We employ MestReNova \cite{mnova} to simulate $^1$H-, $^{13}$C- and HSQC-NMR spectra. The spectra were simulated using deuterated Chloroform as solvent. Default settings were used for all simulations. For $^{13}$C-NMR spectra, $^1$H decoupled spectra were generated. 

We utilize the in-built spectral analysis tools of MestreNova to annotate the spectra. For $^1$H-NMR spectra, we employ the automultiplet analysis function yielding a set of peaks, the type of each peak e.g. doublet, triplet, etc., and the normalized integration of the peak. The same method yields the position and intensity of the peaks in the $^{13}$C-NMR spectra. Similarly, we obtain the position and integration for peaks in the HSCQC spectra. 

\textbf{IR Simulations:} IR spectra can be simulated either by approximating the bonds in the molecule as harmonic oscillators and calculating their frequencies or by measuring the dipole-dipole moment of the molecule over time \cite{esch2021quantitative, mcgill2021predicting}. While the first approach is computationally cheaper it only yields the position and intensity of each peak in the spectrum which can subsequently be broadened e.g. via a Gaussian function. Overtones and anharmonicities are neglected by this approach. On the other hand, a simulated IR spectrum derived from dipole-dipole data does contain these features at the expense of higher computation requirements.

We developed a high throughput pipeline to orchestrate molecular dynamics simulations and calculate the spectra from the molecule's dipole moment. 
Based on a molecule as a SMILES string we generate the corresponding Protein Data Bank (PDB) file and optimize the geometry of the molecule with the General AMBER Force Field (GAFF) \cite{wang2004development}. We choose the same force field for the molecular dynamics simulation and generate the input files for a Large-scale Atomic-Molecular Massively Parallel Simulator (LAMMPS) \cite{thompson2022lammps} simulation using AMBER tools \cite{case2023ambertools}. The system is allowed to equilibrate for 250 ns, before recording the dipole moment of the molecule for a further 250 ns. IR spectra are calculated from the dipole moment according to Braun \cite{braun_calculating_2016}. The simulated spectra have a range from 400--4000 $cm^{-1}$ with a resolution of 2 $cm^{-1}$.

\textbf{MS/MS Simulations:} The development MS/MS simulation tools is advancing rapidly, with new tools and approaches emerging frequently. To capture the current state of the art, we selected three distinct methods that represent different approaches to MS/MS simulation: Competitive Fragmentation Modeling for Metabolite Identification 4.0 (CFM-ID 4.0) \cite{wang2021cfm}, Subformula Classification for Autoregressively Reconstructing Fragmentations (SCARF) \citet{goldman_prefix-tree_2023}, and ICEBERG \citet{goldman_generating_2024}. While SCARF and ICEBERG employ pure machine learning approaches, CFM-ID represents a hybrid methodology combining machine learning with rule-based systems. Important to note is that we use the publically available checkpoints for both SCARF and ICEBERG. The performance report in \citet{goldman_prefix-tree_2023} and \citet{goldman_generating_2024} was obtained using the closed source NIST20 database.

We simulate positive mode Electrospray Ionisation (ESI) MS/MS spectra using hydrogen adducts using all three methods. Additionally, we use CFM-ID to simulate negative mode spectra, generating spectra at three ionization energies (10eV, 20eV, and 40eV) in both positive and negative modes. All three tools provide chemical formula annotations for the fragments in their simulated MS/MS spectra.

\subsection{Experimental vs Simulated Spectra}
To evaluate the similarity of the simulated spectra to experimental ones, we compared them with a set of 251 molecules and their corresponding experimentally measured spectra from \citet{van2023spectroscopy}. Out of these 251 molecules, 96 had all spectroscopic techniques measured and were also simulated in the dataset introduced in this manuscript (excluding HSQC-NMR). Since each spectral technique has a different representation, multiple approaches were required for comparison. Table \ref{tab:exp_vs_sim_tab} presents the spectral similarity between the real-world measured spectra and our simulated approach. Additionally, as a comparison metric, the similarity of each the 96 simulated spectra versus all experimental spectra was calculated. The similarity metrics used are listed in the table (their definitions can be found in Appendix section \ref{sec:A_similiarity}). 

\begin{figure}[t!]
    \centering
    \includegraphics[width=0.9\linewidth]{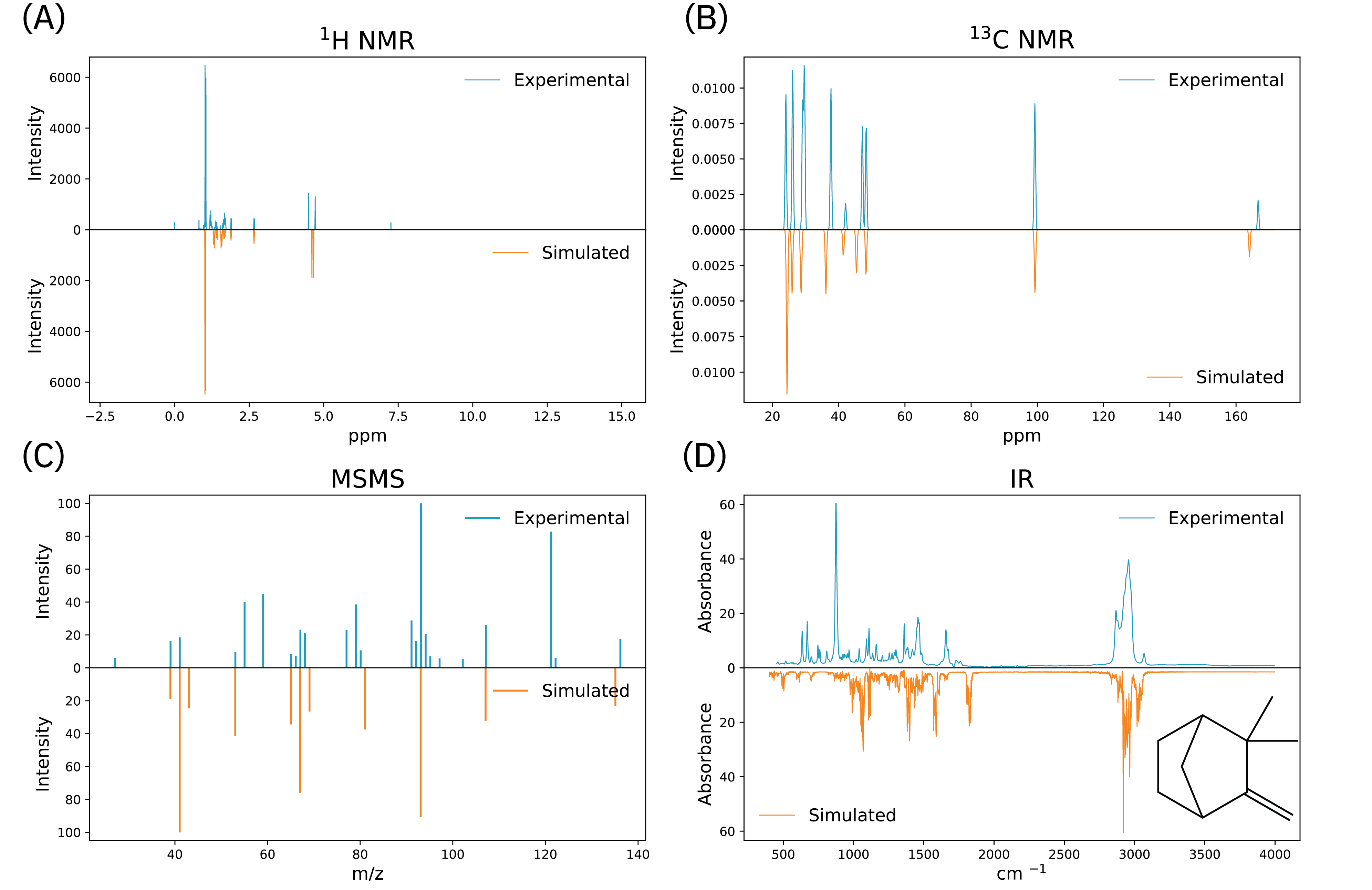}
    \caption{Simulated vs experimentally measured (A) $^1$H-NMR (B) $^{13}$C-NMR (C) MS/MS [+40 eV] and (D) IR  of the molecule 2,2-dimethyl-3-methylenebicyclo[2.2.1]heptane (shown in the lower right).}
    \label{fig:sim_vs_exp}
\end{figure}

\begin{table}[t!]
    \centering
    \setlength\tabcolsep{1pt}
    \renewcommand{\arraystretch}{1.3}
    
    \caption{Similarity metrics between experimental and simulated spectra. For the MS/MS only positive modes are compared as the experimental spectra were measured in this mode. }
    \vspace{5px}
    \begin{tabular}{lccl}
    \toprule
         Spectrum&  Sim. vs Exp.& Sim vs other Exp.&Similarity Metrics\\
         \midrule
         IR&  31.5&  26.6&Cosine Similarity\\ \midrule
MS/MS (CFM-ID, Positive)  [10 eV]& 30.1& 17.9&CosineGreedy Similarity \cite{huber2020matchms}\\
MS/MS (CFM-ID, Positive) [20 eV]& 40.1& 22.6&CosineGreedy Similarity \cite{huber2020matchms}\\
MS/MS (CFM-ID, Positive) [40 eV]& 48.9& 26.6&CosineGreedy Similarity \cite{huber2020matchms}\\
\midrule

MS/MS (SCARF, Positive)& 14.1 & 8.2 &CosineGreedy Similarity \cite{huber2020matchms}\\
MS/MS (ICEBERG, Positive)& 17.0 & 10.9 &CosineGreedy Similarity \cite{huber2020matchms}\\
\midrule

$^1$H-NMR& 21.9& 6.8&Cosine Similarity\\
$^{13}$C-NMR& 48.4& 8.6&CosineGreedy Similarity \cite{huber2020matchms}\\
\bottomrule
    \end{tabular}

    \label{tab:exp_vs_sim_tab}
\end{table}

Relying primarily on cosine similarity-based metrics has a significant limitation. For example, in the case of NMR spectra, even if the shape and integral are accurately simulated, a slight peak shift compared to the experimentally measured spectrum (a common effect caused by the solvent \cite{buckingham1960solvent}) can result in a drastic reduction in cosine similarity. While a chemist would consider the two compared spectra similar, the cosine similarity score would be substantially lower than when the peaks are aligned.
Despite this limitation, it can be seen that, on average, all simulated spectra have a higher similarity to their corresponding experimental spectra compared to average similarity against all other experimental spectra. This shows that the simulated data represents somewhat realistically experimentally measured spectra.
For visual inspection of the similarity between simulated and experimental spectra see Figure \ref{fig:sim_vs_exp}. A more in-depth analysis of the similarity between experimental and simulated spectra on larger datasets is presented in appendix section \ref{app:simtoreal}.

\section{Benchmarks}
\begin{figure}[h!]
    \centering

    \includegraphics[width=\linewidth]{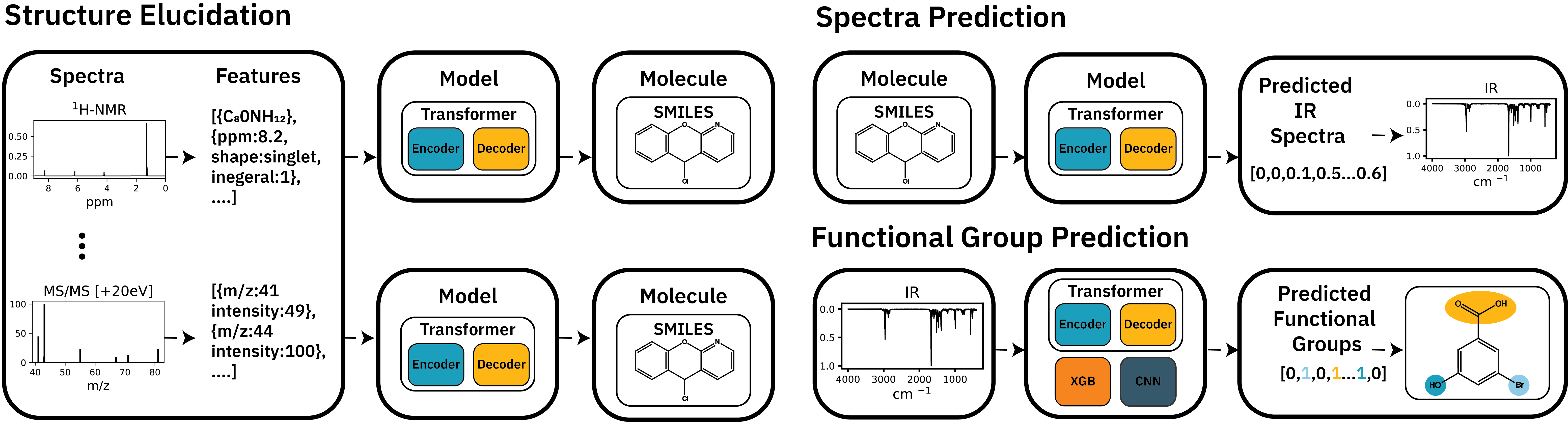}
    \caption{Overview of the benchmarks. Left: Structure elucidation using transformer models. We convert each spectra into a structured text representation to make it ingestible by the model. Top Right: Generation of spectra from molecules using a transformer model. We reuse the same structured text representation. Bottom right: Predicting functional groups from the spectra as a multilabel multiclass classification problem. We assess transformers, a 1D-CNN and gradient boosted trees.}
    \label{fig:benchmark}
\end{figure}

In the following we will present benchmarks on predicting the correct structure, functional groups contained in a molecule and generating spectra from a given molecule. We only evaluate performance on single modalities and leave exploring multimodal tasks for future work. All experiments are conducted with five fold cross validation. An overview of the different tasks is shown in Figure \ref{fig:benchmark}.

\subsection{Structure Elucidation from Spectra}
As described in the introduction, we envision full structure elucidation from spectra as the primary use case for this dataset. To this end, we provide baseline results on predicting the exact chemical structure from spectra (see Table \ref{tab:spec_to_struc}). We train a vanilla encoder-decoder transformer model \cite{vaswani2017attention, klein_opennmt_2017} on each individual modality and on the combination of $^1$H- and $^{13}$C-NMR. More information on the exact model and parameters used can be found in Appendix section \ref{app:model}. In addition to the spectra, we provide the models with the chemical formula, i.e. the elements present in the molecule, as a prior. The chemical formula can be obtained experimentally via high resolution MS.

\begin{table}[b]
    \renewcommand{\arraystretch}{1.3}
    \caption{
        Top--1, Top--5, and Top--10 Accuracy (see Appendix \ref{app:metric}) of a Transformer model trained to predict the chemical structure (SMILES) from the different modalities.
    }
    \label{tab:spec_to_struc}
    \begin{center}
    \begin{tabular}{lccc}
    \toprule
         & Top--1\% & Top--5\% & Top--10\% \\
        \midrule
        IR & 9.97 $\pm$ 0.46 & 21.23 $\pm$ 0.33 & 24.01 $\pm$ 0.42 \\ \midrule
        MS/MS (CFM-ID, Negative) & 20.98 $\pm$ 0.23 & 39.32 $\pm$ 0.19 & 44.93 $\pm$ 0.29 \\
        MS/MS (CFM-ID, Positive) & 23.53 $\pm$ 0.21 & 42.59 $\pm$ 0.14 & 47.53 $\pm$ 0.31 \\
        \midrule
        MS/MS (SCARF, Positive) & 1.92 $\pm$ 0.11 & 5.26 $\pm$ 0.37 & 6.81 $\pm$ 0.48 \\ 
        MS/MS (ICEBERG, Positive) & 15.52 $\pm$ 2.10 & 31.46 $\pm$ 3.28 & 36.22 $\pm$ 3.45 \\
        \midrule
        $^{13}$C-NMR & 51.95 $\pm$ 0.29 & 70.01 $\pm$ 0.21 & 74.12 $\pm$ 0.30 \\
        $^{1}$H-NMR & 64.99 $\pm$ 0.31 & 81.94 $\pm$ 0.31 & 84.07 $\pm$ 0.32 \\ 
        $^{1}$H-NMR + $^{13}$C-NMR & 73.38 $\pm$ 0.08 & 87.94 $\pm$ 0.14 & 89.98 $\pm$ 0.16 \\
        \bottomrule
    \end{tabular}
    \end{center}
\end{table}

To train a transformer model on the spectra we convert the spectra into a structured text representation. For IR and NMR spectra we follow the representations described in earlier works by \citet{alberts2023learning, alberts2023leveraging}. For IR spectra this representation converts the spectrum to a set of 400 tokens each sampled from a fixed position in the spectrum and bins the intensities to tokens. For $^{13}$C-NMR spectra the representation provides the model with the position of each peak in the spectrum, whereas $^{1}$H-NMR spectra we provide the integration and type of each peak in addition its beginning and end. We train a model on the spectra generated by SCARF, ICEBERG and CFM-ID. In addition, for CFM-ID we train a model both for the positive as well as the negative MS/MS spectra. Each peak in a spectra is described using the peaks m/z and intensity. Examples for all representations can be found in Appendix section \ref{app:representations}.

We observe the worst performance for models solely trained on IR spectra followed by MS/MS spectra. On the other hand, $^1$H- and $^{13}$C-NMR perform relatively well. Encouragingly the combination of both $^1$H- and $^{13}$C-NMR performs the best. These results can be explained by the information contained in each modality. While IR spectra can be leveraged easily to determine the functional groups present in a molecule, for larger and more complex molecules the peaks in the spectrum start to overlap rendering it difficult to extract information. The low performance on MS/MS spectra is caused by similar factors: The more complex the molecule, the larger the number of potential fragmentations, increasing the difficulty of assigning a definite structure. Models trained on $^1$H-NMR spectra perform better than $^{13}$C-NMR as $^1$H-NMR spectra typically contain more information. However, as the two types of spectra probe different aspects of the molecule they complement each other resulting in a performance increase of 7.8\% when combined. We also conducted zero-shot experiments of the models trained on simulated data on the \citet{van2023spectroscopy} dataset. These results are shown in appendix section \ref{app:zeroshot}.

\label{sec:struc_elucidation}

\subsection{Functional Group prediction}
\label{sec:func_group}

\begin{table}[b]
    \renewcommand{\arraystretch}{1.3}
    
    \centering
    \caption{F1 scores for predicting functional groups from the different spectra.}
    \vspace{5px}
    \begin{tabular}{lccc}
    \toprule
        Spectrum & XGBoost & 1D-CNN \cite{jung2023automatic} & Transformer \\
        \midrule
         
        IR & 0.834 $\pm$ 0.001 & \textbf{0.895 $\pm$ 0.002} & 0.881 $\pm$ 0.021  \\ \midrule
        MS/MS (CFM-ID, Positive) & 0.725 $\pm$ 0.002 & 0.645 $\pm$ 0.001 & \textbf{0.897 $\pm$ 0.012}  \\
        MS/MS (CFM-ID, Negative) & 0.761 $\pm$ 0.001 & 0.648 $\pm$ 0.006 & \textbf{0.905 $\pm$ 0.009} \\ \midrule
        MS/MS (SCARF, Positive) & 0.763 $\pm$ 0.001 & 0.737 $\pm$ 0.003 & \textbf{0.771 $\pm$ 0.004} \\
        MS/MS (ICEBERG, Positive) & 0.734 $\pm$ 0.001 & 0.677 $\pm$ 0.001  & \textbf{0.885 $\pm$ 0.007} \\ \midrule
        $^{13}$C-NMR & 0.804 $\pm$ 0.001 & 0.674 $\pm$ 0.056 & \textbf{0.913 $\pm$ 0.017} \\ 
        $^{1}$H-NMR & 0.797 $\pm$ 0.003 & 0.839 $\pm$ 0.005 & \textbf{0.935 $\pm$ 0.031} \\ 

    \bottomrule
    \end{tabular}
    \label{tab:func_group2}
\end{table}

Another task that can be explored using the dataset is predicting the functional groups present in the structure from the spectra. We extract functional groups from the molecules using the SMARTS \cite{ehrt2020smarts} pattern defined in \ref{sec:A_functional_group}. While not as useful to chemists as full structure elucidation, the success of a chemical reaction can in most cases be determined by a change in functional groups. We approach this task as a multiclass, multilabel classification problem. As such we evaluate the performance of three different models, a boosted tree classifier  \cite{chen_xgboost_2016}, a 1D-CNN as implemented by \citet{jung2023automatic} and a transformer model, in predicting  the functional groups present in the target molecule. The performance of the models on the modalities is shown in Table \ref{tab:func_group2}. We train the boosted gradient tree and 1D-CNN on the non processed vector form of each spectrum. In contrast we employ the same representations as used in for structure elucidation task for the transformer model. Unlike the previous task,  we do not include the chemical formula as an input. 

Across four modalities the transformer trained on the structured text representations outperforms both the 1D-CNN and the gradient boosted trees. Only on IR spectra is the performance of the 1D-CNN marginally better than the Transformer model. In contrast to MS/MS and the NMR spectra, IR spectra are not sparse explaining the good performance of the 1D-CNN.

\subsection{Spectra prediction}
We primarily conceived the dataset to explore structure elucidation. However, the dataset can also be used for the reverse, i.e. predicting the corresponding spectrum given a target molecule. To this end we train a transformer model to predict the from the molecule for each modality. We use the same model architecture and representations as in section \ref{sec:struc_elucidation}. This mean that while we predict the whole spectrum for IR spectra, for all other modalities a processed form of the spectrum is generated. In the case of the MS/MS spectra this consists of the m/z of each peak and it's intensity. Similarly for $^{13}$C-NMR spectra the model predicts the position of each peak. However, for $^{1}$H-NMR spectra we predict the start and end of each peak, it's type and integration.

To compare the predicted and target spectrum we use two similarity metrics: One one hand we employ greedy cosine similarity and on the other the exact token accuracy. For MS/MS, $^{13}$C- and $^{1}$H-NMR spectra we compute the cosine similarity by first aligning the peaks in the predicted and target spectrum before calculating the similarity. The results are shown in Table \ref{tab:pred_spectra}.

For IR spectra we observe both a low cosine and token similarity. This is likely caused by the represenation used for predicting the spectra as a sequence of 400 tokens has to be generated for each spectrum. Other approaches such as graph neural networks as proposed by \citet{mcgill2021predicting} may show better performance. For both positive and negative MS/MS we observe a decrease in performance with an increase in the ionisation energy, likely a result of molecules fragmenting to a larger extent at higher ionisation energy and as such resulting in a more complex spectrum. Predicted $^{1}$H- and $^{13}$C-NMR spectra both exhibit a high cosine similarity while showing a small token accuracy. This is caused by two factors: One one hand only the position of the peak is used to calculate the similarity and on the other hand the token accuracy requires an exact match of the token, i.e. even if the predicted peak has an error of only 0.1ppm it would be deemed false.

\subsection{Other tasks to explore}

\begin{table}
    \centering
    \renewcommand{\arraystretch}{1.3}
    
    \caption{Cosine similarity and token accuracy of transformer models when predicting spectra from structure. We predict an individual MS/MS spectra for each ionisation energy}
    \vspace{5px}
    \begin{tabular}{lcc}
    \toprule
        Spectrum&  Cosine Similarity & Token Accuracy\\
        \midrule
        IR&  23.91 $\pm$ 0.14 &  13.55 $\pm$ 0.16\\ \midrule
        MS/MS (Positive)   [10 eV] & 83.94 $\pm$ 0.10 & 31.58 $\pm$ 0.09 \\
        MS/MS (Positive)   [20 eV] & 77.09 $\pm$ 0.18 & 11.05 $\pm$ 0.13 \\
        MS/MS (Positive)   [40 eV] & 66.35 $\pm$ 0.15 & 6.94 $\pm$ 0.16 \\ 
        \midrule
        MS/MS (Negative) [10 eV] & 82.87 $\pm$ 0.25 & 33.92 $\pm$ 0.19\\
        MS/MS (Negative)  [20 eV] & 75.86 $\pm$ 0.18 & 11.82 $\pm$ 0.11 \\
        MS/MS (Negative)   [40 eV] & 69.50 $\pm$ 0.23 & 8.95 $\pm$ 0.17 \\
        \midrule
        MS/MS (SCARF, Positive) & 66.39 $\pm$ 0.03 & 5.04 $\pm$ 0.08 \\
        MS/MS (ICEBERG, Positive) & 63.17 $\pm$ 0.01 & 4.62 $\pm$ 0.04 \\
        \midrule
        $^{13}$C-NMR & 92.69 $\pm$ 0.31 &  35.7 $\pm$ 0.27\\ \midrule
        $^{1}$H-NMR & 94.86 $\pm$ 0.29 &  17.93 $\pm$ 0.24\\ 

    \bottomrule
    \end{tabular}

    \label{tab:pred_spectra}
\end{table}

While we benchmark three different tasks that could be of interest to researchers, the dataset can also be used for various other ML applications. The following ideas serve as starting points for potential research opportunities to explore.

\textbf{Including more information:} Typically, when predicting the structure of molecules from spectra, chemists do not start from scratch. Instead, they leverage prior knowledge of the reaction performed to make informed initial guesses about the structure. These guesses may include the desired product, the starting material, or plausible side products. Chemists then use clues obtained from various spectra to confirm or eliminate these initial hypotheses. In this context, we propose a task where a model is provided with a set of molecules and a spectrum, and it predicts which molecule from the set corresponds to the given spectrum.

\textbf{Mixtures:} While the previous sections discussed structure elucidation from pure compounds, in reality chemists typically encounter mixtures far more often than pure compounds, e.g. a reaction is a mixture of a set of compounds. The mixtures commonly need to be separated into their constituent components before definite structure elucidation can be carried out. If the components of a mixture could identified accurately this would greatly aid chemists. The spectra of mixtures can be constructed as convex combinations of their constituent components for NMR and IR spectra. As such this dataset could be used to construct the spectra of complex mixture based on which the components of the mixture can be predicted.

\textbf{Representations:} While our study utilized SMILES as the primary form to represent molecules there are a variety of different chemical representations that could be explored, reaching from SELFIES\cite{krenn2020self}, deep-SMILES\cite{o2018deepsmiles} as well as generating graph instead of a text-based representation.  Additionally, the spectral information could be represented in various ways. For instance, generating figures from vector representations and employing image-based models.

\textbf{Multimodal Approaches:} Combining the different types of spectra for a true multimodal model is significantly harder than e.g. predicting the structure from a single modality. Not only does the best representation for each modality need to be considered but also how to combine the different modalities. We suggest the following approach: First optimise the representation for each modality individually before considering how to combine them. Here the approaches range from early fusion, in practice often a simple concatenation of the different modalities, over medium fusion, embed each modality then fuse, to late fusion. Another consideration is how to process each modality and weigh each modality. Each of these factors would need to be considered when building a multimodal model.

\section{Conclusion and Limitations}
\label{sec:limitations}
In this study we introduce the first multimodal spectroscopic dataset for structure elucidation including six different types of spectra for 790k molecules sampled from USPTO. We conduct a series of experiments on structure elucidation, generating a spectrum from a molecule and evaluate boosted gradient trees, 1D-CNNs and transformer models on classifying which functional groups are present in a molecule based on the spectra. 

However, the dataset also has multiple limitations, the largest being that all data is simulated. As a result, there is a distribution shift between simulated and experimental spectra and models trained on the data in this set will likely benefit from further finetuning on experimental spectra. The efficacy of these models is inherently determined by the fidelity of the underlying simulation used to generate the spectra. Another limitation is the chemical space covered by USPTO. The USPTO dataset contains molecules sourced from patents and as a result, is biased towards synthesisable molecules with applications in industry. This means that models trained on our dataset may not perform as well on molecules outside of this scope, e.g. natural products.

We hope that this work can address the severe lack of openly available spectroscopic datasets and serve as a foundation to build models capable of automated structure elucidation for chemistry and with that streamline the molecular discovery pipeline from synthesis to structure determination.

\begin{ack}
This publication was created as part of NCCR Catalysis (grant number 180544), a National Centre of Competence in Research funded by the Swiss National Science Foundation.
\end{ack}
\section*{Code and Data availability}
The code for the benchmark and the dataset used in this study is publicly available under the following link: \url{https://github.com/rxn4chemistry/multimodal-spectroscopic-dataset}.

\bibliographystyle{unsrtnat}
\bibliography{neurips_data_2023.bib}

\newpage
\setcounter{section}{0}
\renewcommand*{\thesection}{\Alph{section}}
\newpage

\section{Appendix}

\subsection{Sample of the dataset}
\label{app:sample_spectra}

\begin{figure}[h!]
    \centering
    \includegraphics[width=1.0\textwidth]{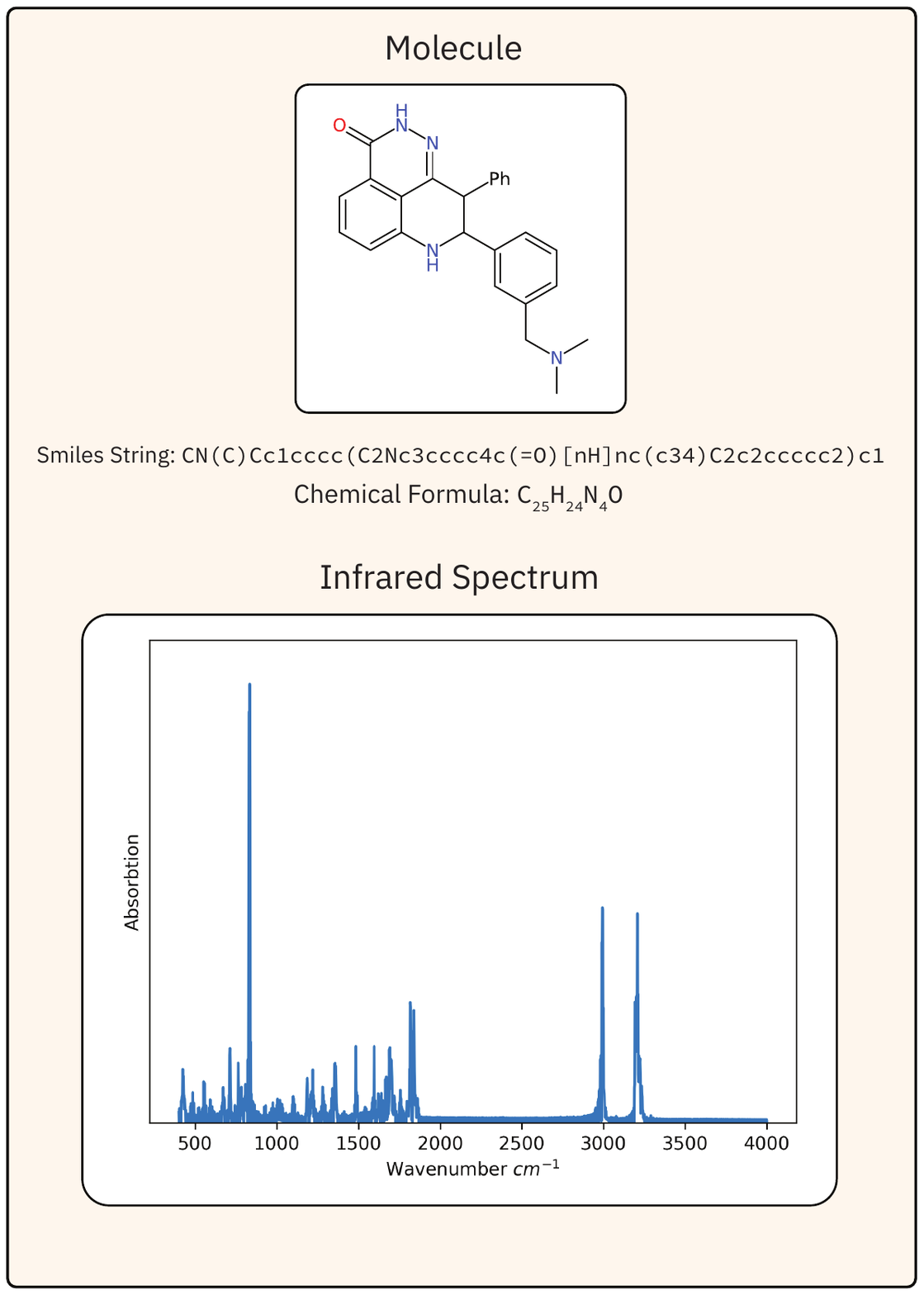}
\end{figure}

\newpage

\begin{figure}[h!]
    \centering
    \includegraphics[width=1.0\textwidth]{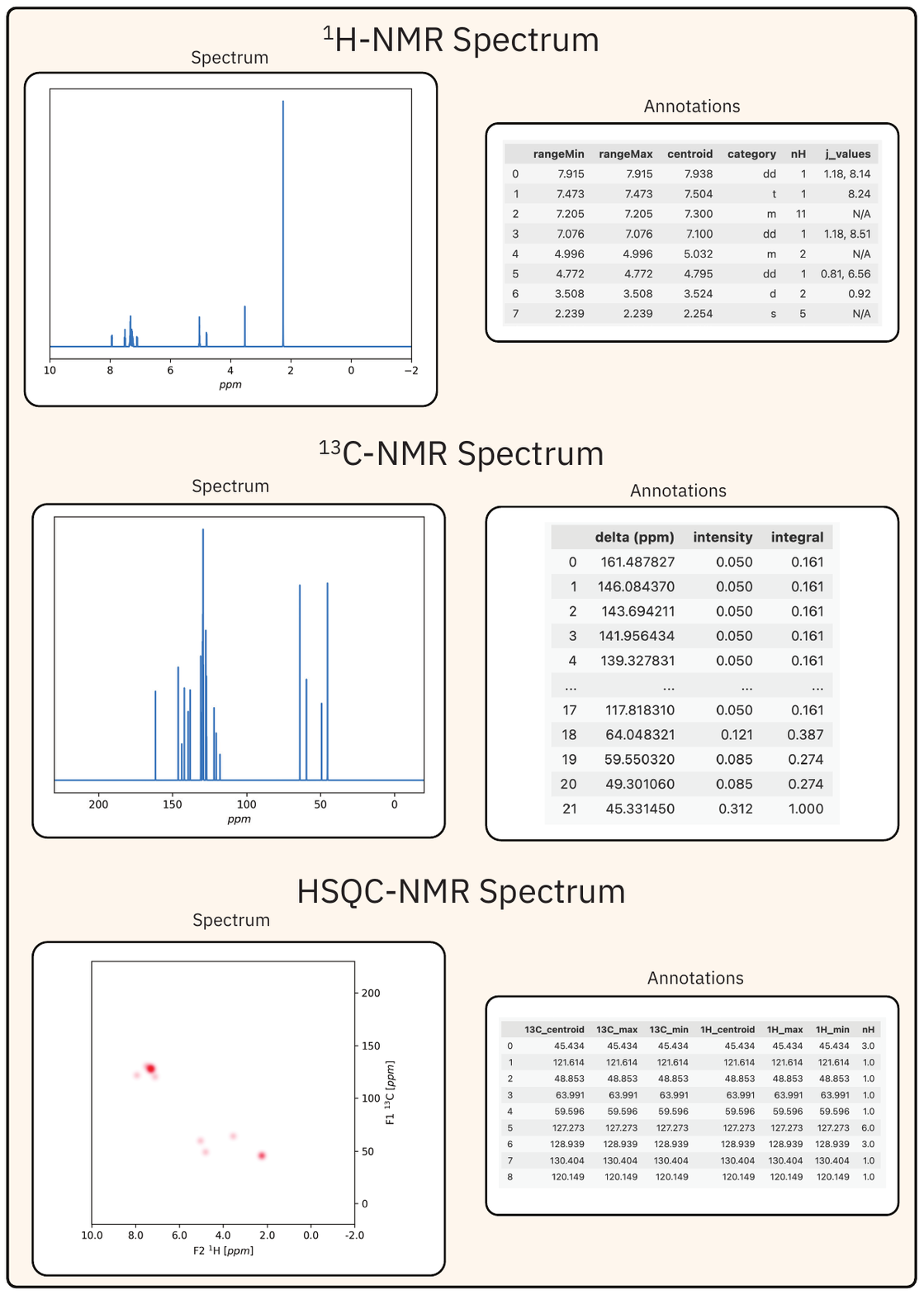}
\end{figure}

\newpage

\begin{figure}[h!]
    \centering
    \includegraphics[width=1.0\textwidth]{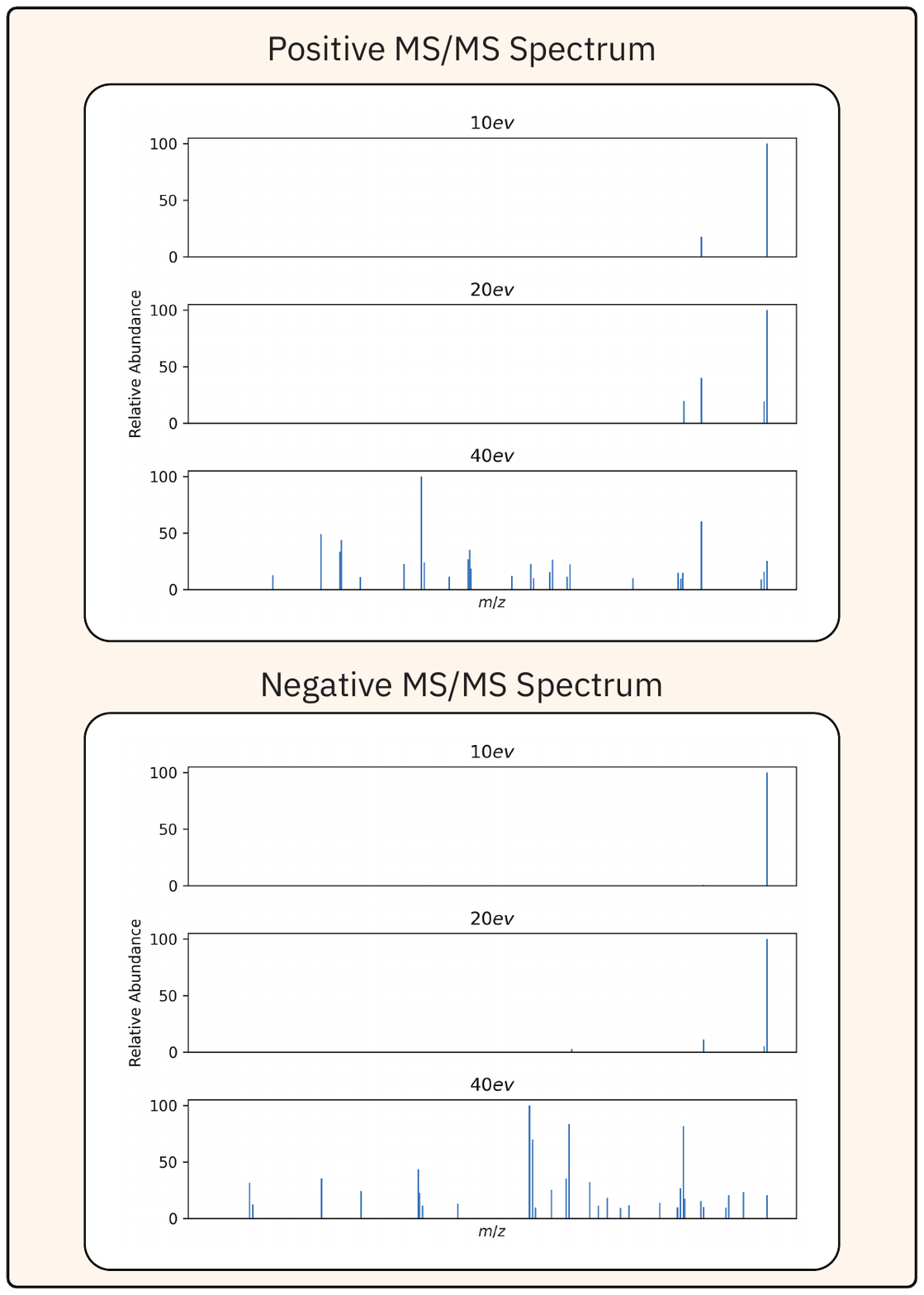}
\end{figure}

\newpage

\begin{figure}[h!]
    \centering
    \includegraphics[width=1.0\textwidth]{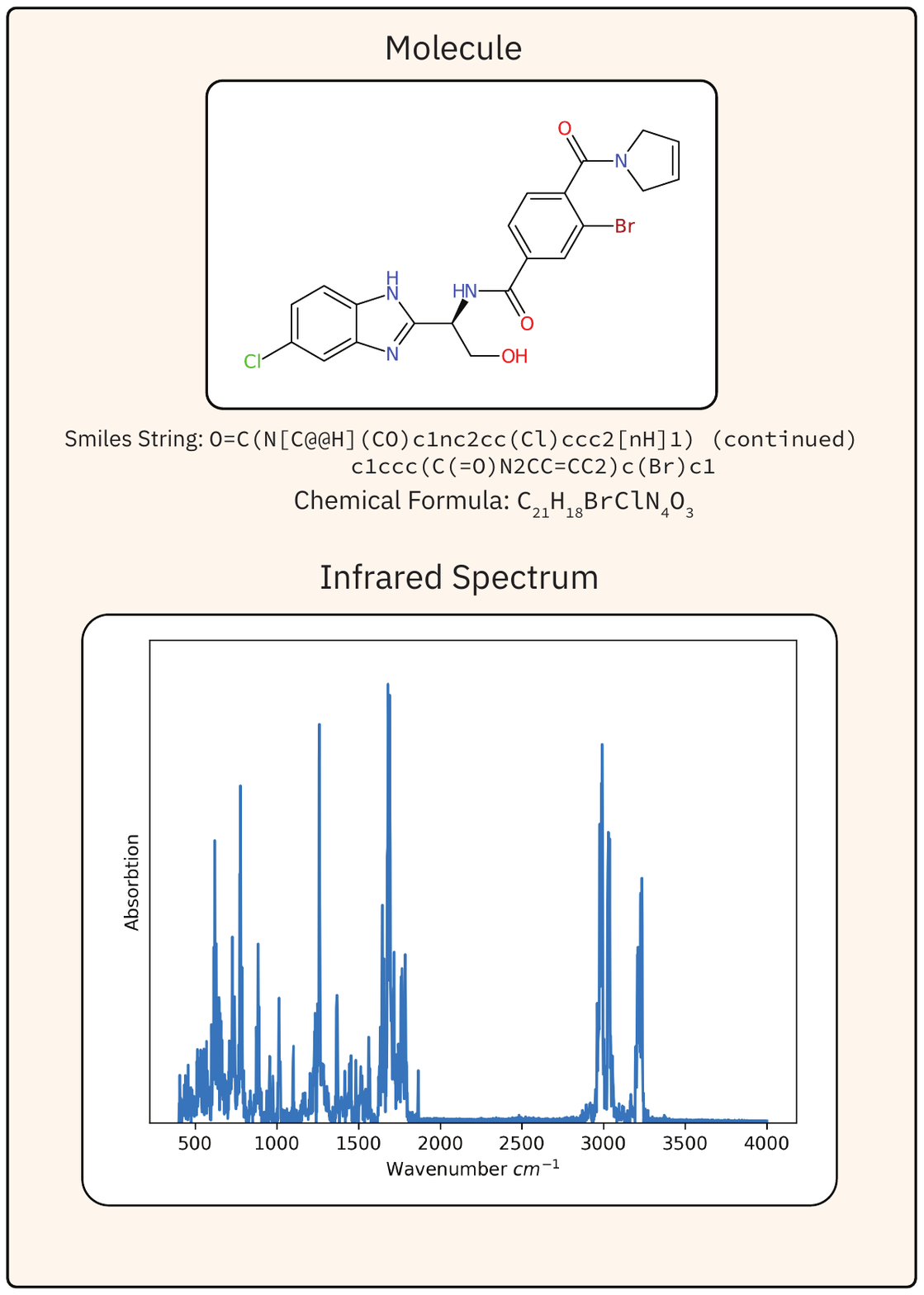}
\end{figure}

\newpage

\begin{figure}[h!]
    \centering
    \includegraphics[width=1.0\textwidth]{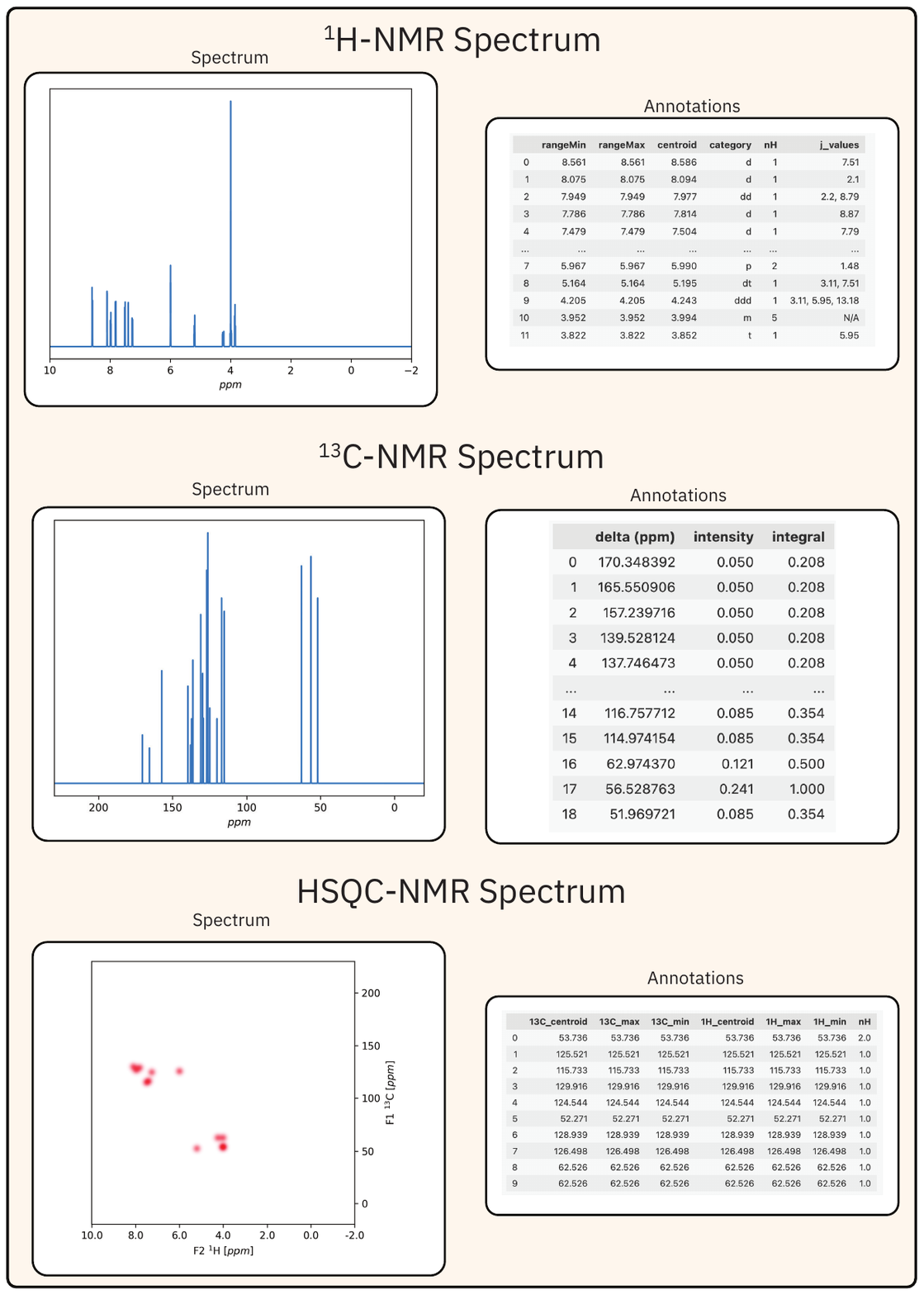}
\end{figure}

\newpage

\begin{figure}[h!]
    \centering
    \includegraphics[width=1.0\textwidth]{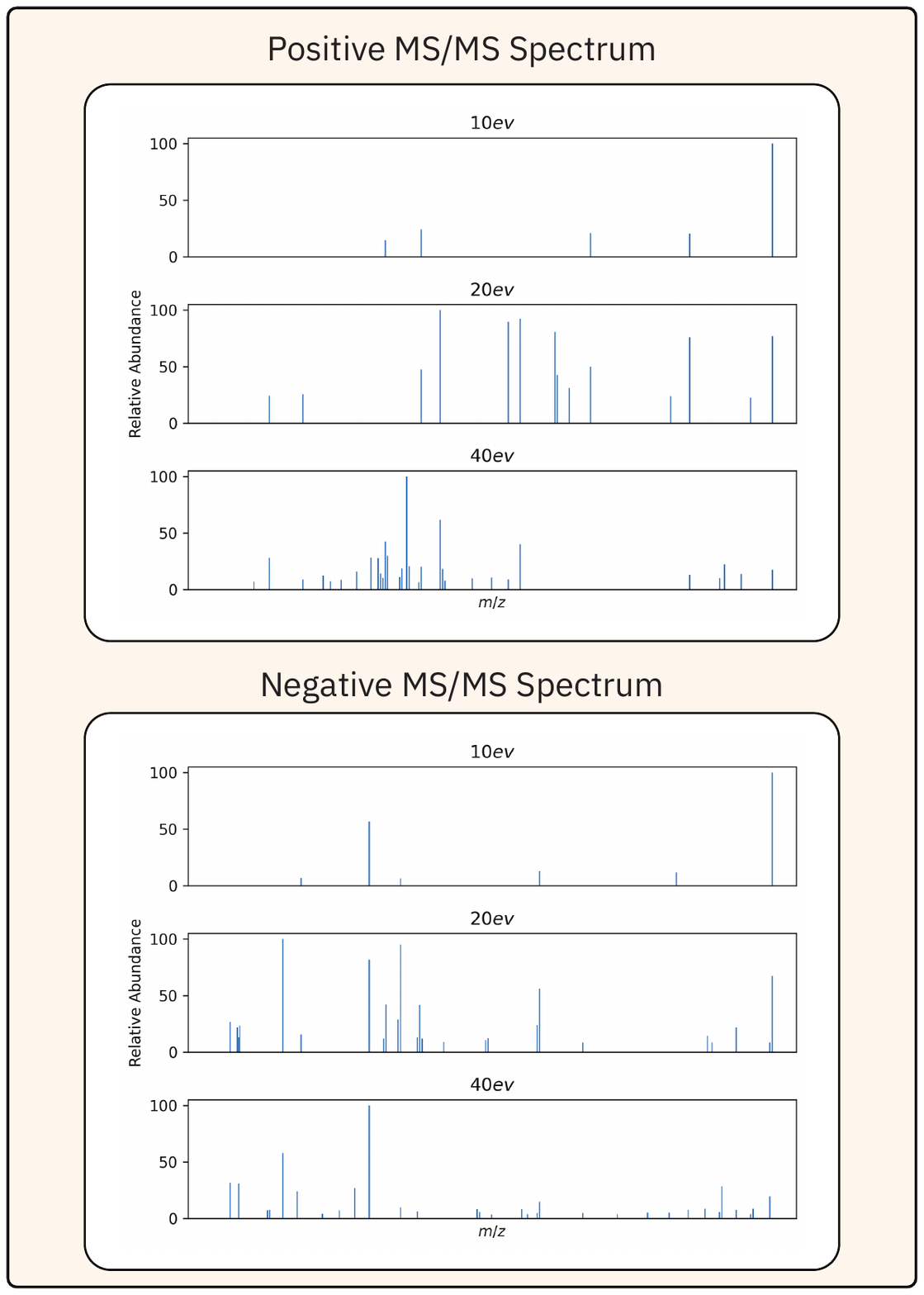}
\end{figure}

\newpage

\subsection{Functional Group Analysis}
The count of functional groups in the molecular structures was analyzed by SMARTS \cite{smarts} pattern matching using RdKit \cite{rdkit}. The applied SMARTS patterns were derived from \citet{jung2023automatic} and listed in Table \ref{tab:A_smarts_pattern}. Each molecule can have only one occurrence of a particular functional group. For instance, if a molecule has two alcohol groups, the occurrence counts only once. However, each molecule can have multiple occurrences of different functional groups.

\begin{table}[h]
\caption{SMARTS pattern used for functional group analysis.}
\begin{tabular}{ll}
\hline
Functional Group & SMARTS Pattern \\
\hline
Acid anhydride & $\left[\mathrm{CX3}\right]\left(=\left[\mathrm{OX1}\right]\right)\left[\mathrm{OX2}\right]\left[\mathrm{CX3}\right]\left(=\left[\mathrm{OX1}\right]\right)$ \\
Acyl halide & $\left[\mathrm{CX3}\right]\left(=\left[\mathrm{OX1}\right]\right)\left[\mathrm{F,Cl,Br,I}\right]$ \\
Alcohol & $\left[\#6\right]\left[\mathrm{OX2H}\right]$ \\
Aldehyde & $\left[\mathrm{CX3H1}\right]\left(=\mathrm{O}\right)\left[\#6,\mathrm{H}\right]$ \\
Alkane & $\left[\mathrm{CX4;H3,H2}\right] $\\
Alkene & $\left[\mathrm{CX3}\right]=\left[\mathrm{CX3}\right] $\\
Alkyne &$ \left[\mathrm{CX2}\right]\#\left[\mathrm{CX2}\right]$ \\
Amide & $\left[\mathrm{NX3}\right]\left[\mathrm{CX3}\right]\left(=\left[\mathrm{OX1}\right]\right)\left[\#6\right] $\\
Amine & $\left[\mathrm{NX3;H2,H1,H0;!}\$\left(\mathrm{NC}=\mathrm{O}\right)\right] $\\
Arene & $\left[\mathrm{cX3}\right]1\left[\mathrm{cX3}\right]\left[\mathrm{cX3}\right]\left[\mathrm{cX3}\right]\left[\mathrm{cX3}\right]\left[\mathrm{cX3}\right]1 $\\
Azo compound & $\left[\#6\right]\left[\mathrm{NX2}\right]=\left[\mathrm{NX2}\right]\left[\#6\right] $\\
Carbamate & $\left[\mathrm{NX3}\right]\left[\mathrm{CX3}\right]\left(=\left[\mathrm{OX1}\right]\right)\left[\mathrm{OX2H0}\right]$ \\
Carboxylic acid & $\left[\mathrm{CX3}\right]\left(=\mathrm{O}\right)\left[\mathrm{OX2H}\right]$ \\
Enamine & $\left[\mathrm{NX3}\right]\left[\mathrm{CX3}\right]=\left[\mathrm{CX3}\right] $\\
Enol & $\left[\mathrm{OX2H}\right]\left[\#6\mathrm{X3}\right]=\left[\#6\right]$ \\
Ester & $\left[\#6\right]\left[\mathrm{CX3}\right]\left(=\mathrm{O}\right)\left[\mathrm{OX2H0}\right]\left[\#6\right]$ \\
Ether & $\left[\mathrm{OD2}\right]\left(\left[\#6\right]\right)\left[\#6\right] $\\
Haloalkane & $\left[\#6\right]\left[\mathrm{F,Cl,Br,I}\right] $\\
Hydrazine & $\left[\mathrm{NX3}\right]\left[\mathrm{NX3}\right] $\\
Hydrazone & $\left[\mathrm{NX3}\right]\left[\mathrm{NX2}\right]=\left[\#6\right] $\\
Imide & $\left[\mathrm{CX3}\right]\left(=\left[\mathrm{OX1}\right]\right)\left[\mathrm{NX3}\right]\left[\mathrm{CX3}\right]\left(=\left[\mathrm{OX1}\right]\right) $\\
Imine & $\left[\$\left(\left[\mathrm{CX3}\right]\left(\left[\#6\right]\right)\left[\#6\right]\right),\$\left(\left[\mathrm{CX3H}\right]\left[\#6\right]\right)\right]=\left[\$\left(\left[\mathrm{NX2}\right]\left[\#6\right]\right),\$\left(\left[\mathrm{NX2H}\right]\right)\right] $\\
Isocyanate & $\left[\mathrm{NX2}\right]=\left[\mathrm{C}\right]=\left[\mathrm{O}\right] $\\
Isothiocyanate & $\left[\mathrm{NX2}\right]=\left[\mathrm{C}\right]=\left[\mathrm{S}\right] $\\
Ketone & $\left[\#6\right]\left[\mathrm{CX3}\right]\left(=\mathrm{O}\right)\left[\#6\right]$ \\
Nitrile & $\left[\mathrm{NX1}\right]\#\left[\mathrm{CX2}\right] $\\
 Phenol & $\left[\mathrm{OX2H}\right]\left[\mathrm{cX3}\right]:\left[\mathrm{c}\right] $\\
Phosphine & $\left[\mathrm{PX3}\right] $\\
Sulfide & $\left[\#16\mathrm{X2H0}\right] $\\
Sulfonamide &$ \left[\#16\mathrm{X4}\right]\left(\left[\mathrm{NX3}\right]\right)\left(=\left[\mathrm{OX1}\right]\right)\left(=\left[\mathrm{OX1}\right]\right)\left[\#6\right] $\\
Sulfonate & $\left[\#16\mathrm{X4}\right]\left(=\left[\mathrm{OX1}\right]\right)\left(=\left[\mathrm{OX1}\right]\right)\left(\left[\#6\right]\right)\left[\mathrm{OX2H0}\right]$ \\
Sulfone & $\left[\#16\mathrm{X4}\right]\left(=\left[\mathrm{OX1}\right]\right)\left(=\left[\mathrm{OX1}\right]\right)\left(\left[\#6\right]\right)\left[\#6\right] $\\
Sulfonic acid & $\left[\#16\mathrm{X4}\right]\left(=\left[\mathrm{OX1}\right]\right)\left(=\left[\mathrm{OX1}\right]\right)\left(\left[\#6\right]\right)\left[\mathrm{OX2H}\right]$ \\
Sulfoxide & $\left[\#16\mathrm{X3}\right]=\left[\mathrm{OX1}\right] $\\
Thial & $\left[\mathrm{CX3H1}\right]\left(=\mathrm{S}\right)\left[\#6,\mathrm{H}\right]$ \\
Thioamide & $\left[\mathrm{NX3}\right]\left[\mathrm{CX3}\right]=\left[\mathrm{SX1}\right] $\\
Thiol &$ \left[\#16\mathrm{X2H}\right] $\\
\hline
\end{tabular}
\label{tab:A_smarts_pattern}
\end{table}

\label{sec:A_functional_group}

\subsection{Functional Group influence on $^1$H-NMR spectra}
\label{sec:A_aromatic}
\begin{figure}[h]
    \centering
    \includegraphics[width=0.65\textwidth]{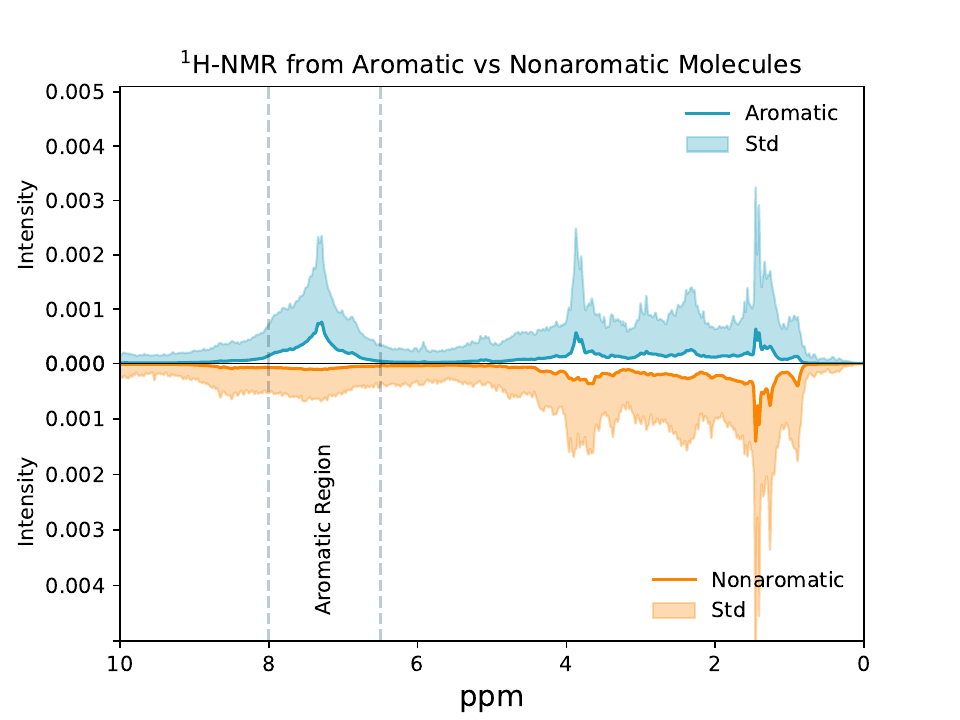}
    \caption{Aromatic molecules $^1$H-NMR spectra averaged and plotted against the average nonaromatic molecules $^1$H-NMR spectra }
    \label{fig:a_aromatic_vs_nonaromatic}
\end{figure}
To investigate if the simulated $^1$H-NMR spectra exhibit the behavior described in the literature of having peaks in the region between 6.0 – 8.7 ppm for aromatic compounds, we divided the spectra into two groups: one containing molecules with aromatic atoms and one with molecules without aromatic atoms. Then, we averaged all the spectra within each group and plotted the averaged spectra with the corresponding standard deviations (see Figure \ref{fig:a_aromatic_vs_nonaromatic}). It can be clearly seen that the non-aromatic molecules lack a signal in the literature-reported aromatic range, as expected. Conversely, the averaged spectra of the aromatic molecules show a distinct peak region in the aromatic range. This indicates that the simulations tend to correctly predict the aromatic regions of the $^1$H-NMR spectra.

\subsection{Metrics}

\textbf{Top--N}
We define accuracy as the exact character match of the predicted vs the true SMILES string. All SMILES in the dataset are canonicalized and in addition, we canonicalize all predicted SMILES strings after generation. Top--N accuracies are calculated as the fraction of samples in which the correct SMILES string is found within the Top--N predictions.

 \label{sec:A_similiarity}
\textbf{Cosine Similarity:}\\
The cosine similarity, $S_C(A, B)$, is defined as the cosine of the angle $\theta$ between two vectors $A$ and $B$, representing the continuous spectra. It is calculated as:
\begin{equation}
S_C(A, B) := \cos(\theta) = \frac{A \cdot B}{|A| |B|}
\end{equation}
\textbf{Chemical Similarity:}\\
The chemical similarity between two molecules is calculated by computing the Tanimoto similarity between their Morgan fingerprints (built using RDKit \cite{rdkit}). The Morgan fingerprints are generated with a length of 1024 and radius of 2. The Tanimoto similarity, $T(A, B)$, between two bit vectors $A$ and $B$, representing the molecular fingerprints, is defined as:
\begin{equation}
T(A, B) = \frac{A \cdot B}{|A|^2 + |B|^2 - A \cdot B}
\end{equation}
\textbf{CosineGreedy:}
The CosineGreedy class from the MatchMS package \cite{huber2020matchms} calculates a modified version of the cosine similarity between two mass spectra. Instead of treating the spectra as vectors across all possible m/z values, it identifies pairs of peaks between the two spectra that have m/z values within a specified tolerance. It then uses a greedy approach to find the best set of matched peak pairs, rather than solving the optimal assignment problem. The cosine similarity score is calculated based on the intensities of these matched peak pairs, optionally weighted by the m/z values raised to a specified power. It can be adapted to a list of NMR peaks by giving instead of the m/z value the ppm, and instead of the intensity giving the integral value. 

\subsection{Sim To Real Gap}
\label{app:simtoreal}

In the following section, we conduct additional experiments to assess the similarity between the computed and experimental spectra. We use larger databases and also measure the similarity between a simulated spectra and the experimental spectra of the molecule with the closest Tanimoto similarity. Below we will outline the experimental datasets used for each modality followed by the similarities presented in Table \ref{tab:sim_exp_v2}. We did not simulate additional spectra but used spectra of molecules contained in both our database and the simulated one.

\begin{itemize}
    \item \textbf{IR} (NIST Gas Phase IR Database \cite{nist}): 2.375 Spectra
    \item \textbf{MS/MS} (GNFPS as prepared by \citet{goldman_prefix-tree_2023}): 640 Spectra 
    \item \textbf{$^{13}$C-NMR} (nmrshiftdb2 \cite{kuhn2024twenty}): 6.627 Spectra 
    \item \textbf{$^1$H-NMR} (Pistachio \cite{noauthor_nextmove_nodate}): We extract the experimental $^1$H-NMR from the patent texts of the entries in Pistachio. In total we compare 10.000 $^1$H-NMR from Pistachio to the text representation (i.e. centroid,  integration and type of each peak) of our simulated $^1$H-NMR spectra
\end{itemize}

\begin{table}[h!]
\footnotesize
\setlength\tabcolsep{4pt}
\renewcommand{\arraystretch}{1.2}
\caption{
Additional similarity experiments: Column \textit{Exp.} compares the similarity of the simulated spectrum to the same experimental one, \textit{Tanimoto} compares the simulated spectrum of the molecule to the experimental spectrum of the molecule with the next closest Tanimoto similarity. \textit{All Others} compares the similarity of a particular simulated spectra to all other experimental spectra.
}
\vspace{5pt}
\label{tab:sim_exp_v2}
\begin{tabular}{llcccc}
\hline
\textbf{Modality} & \textbf{Metric} & \textbf{Exp.} & \textbf{Tanimoto} & \textbf{All Others} & \textbf{Avg. Tanimoto Similarity} \\
\hline
IR & Cosine Similarity & 0.366±0.149 & 0.195±0.141 & 0.190±0.113 & 0.814±0.112 \\
MSMS (CFM-ID, 10 eV) & Greedy Cosine  & 0.486±0.342 & 0.083±0.208 & 0.011±0.009 & 0.714±0.153 \\
MSMS (Scarf) & Greedy Cosine  & 0.148±0.154 & 0.092±0.125 & 0.043±0.019 & 0.714±0.153 \\
MSMS (Iceberg) & Greedy Cosine  & 0.812±0.184 & 0.215±0.265 & 0.044±0.019 & 0.714±0.153 \\
$^1$H-NMR & Greedy Cosine  & 0.941±0.069 & 0.826±0.135 & 0.664±0.094 & 0.687±0.105 \\
$^{13}$C-NMR & Greedy Cosine  & 0.915±0.137 & 0.534±0.218 & 0.175±0.051 & 0.795±0.108 \\
\hline
\end{tabular}
\end{table}

\subsection{Model}
\label{app:model}

For all models, we employ the same 90/10 train test split. The seeds are fixed for all runs.

\subsubsection{Gradient Boosted Tree}

All experiments using gradient boosted trees employ the XGBoost library \cite{DBLP:journals/corr/ChenG16} using the default settings on 32cpu cores:
\begin{verbatim}
    n estimators: 100
    base score: 0.5
    gamma: 0
    learning rate: 0.1 
    max delta step: 0
    max depth: 10
\end{verbatim}

\subsubsection{1D-CNN Model}
We use the implementation by \citet{jung2023automatic} for all 1D-CNN experiments and reuse the same hyperparameters. This 1D-CNN employs three convolutional followed by three fully connected layers. The model is trained with an Adam optimiser for 41 epochs on A100 GPU using Keras \cite{chollet2015keras}. 

\subsubsection{Transformer Model}

We employ a vanilla encoder-decoder transformer as implemented in the OpenNMT-py library \cite{noauthor_opennmt-py_2023, klein_opennmt_2017} with four layers each for the encoder and decoder and a hidden dimension of 512. We train all transformer models for 250k steps amounting to approximately 35h on a A100 GPU. Further hyperparameters can be found below:
\begin{verbatim}
layers: 4
heads: 8
word_vec_size: 512
hidden_size: 512
transformer_ff: 2048
optim: adam
adam_beta1: 0.9
adam_beta2: 0.998
decay_method: noam
learning_rate: 2.0
batch_size: 4096
activation function: ReLu
dropout: 0.1
\end{verbatim}

\subsection{Evaluation metrics}
\label{app:metric}
\textbf{F1 Score }:\\
The F1 score is a measure of a model's performance that combines precision and recall into a single score. It is calculated as the harmonic mean of precision and recall, given by:
\begin{equation}
F1 = 2 \cdot \frac{\text{precision} \cdot \text{recall}}{\text{precision} + \text{recall}}
\end{equation}
where precision is the fraction of true positives among the predicted positives, and recall is the fraction of true positives among the actual positives. The F1 score ranges from 0 to 1, with higher values indicating better performance.

\textbf{Top-k Accuracy}:\\
Top-k Accuracy measures the fraction of instances where the correct answer is among the top-k ranked predictions made by the model. \begin{equation}
\text{Top-}k\text{ Accuracy} = \frac{1}{N} \sum_{i=1}^{N} \mathbb{1}\left(y_i \in \hat{y}_{i}^{(k)}\right)
\end{equation}
N:Total number of samples,$y_i$ True label for the $i\text{-th sample}$
$\hat{y}_{i}^{(k)}$Set of the top  k \text{ predicted labels for the } i\text{-th sample}
$\mathbb{1}(\cdot)$ Indicator function, returns 1 if  $y_i \in \hat{y}_{i}^{(k)}$ otherwise 0

\subsection{Representations for Transformers}
\label{app:representations}

\subsubsection{Molecules}
All molecules in this study are treated as SMILES, a string based molecular representation. We tokenize the molecules following \citet{schwaller2019molecular} using the following regex: 

\begin{verbatim}
(\[[^\]]+]|Br?|Cl?|N|O|S|P|F|I|b|c|n|o|s|p|\(|\)|\.|=|#||\+|\\\\\/|:||@|
\end{verbatim}
\begin{verbatim}
\?|>|\*|\$|\%[0–9]{2}|[0–9])
\end{verbatim}
\subsubsection{Infrared Spectra Representation}

\begin{figure}[h!]
    \centering
    \includegraphics[width=0.88\textwidth]{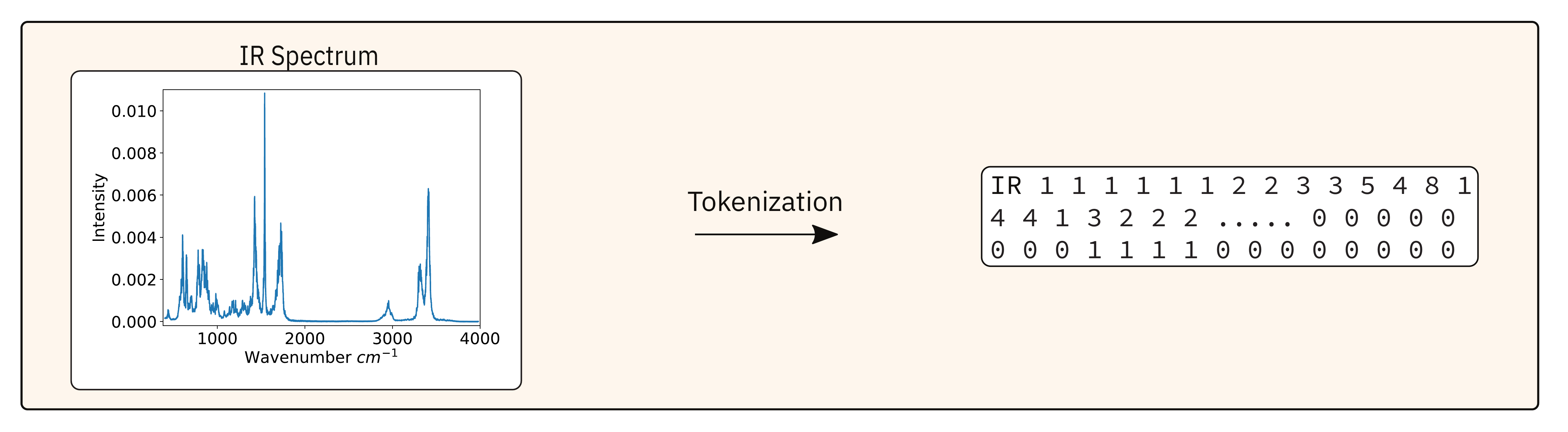}
    \caption{Tokenization of IR spectra.}
    \label{fig:ir_input}
\end{figure}

The IR spectra generated via molecular dynamics range from 400 to 4000cm$^{-1}$ with a resolution of 2cm$^{-1}$, i.e. it is a vector of size 1800. To convert this vector into a structured text representation ingestible by a transformer model we first downsample the spectrum to a vector of size 400 via linear interpolation. We subsequently scale the spectrum to a range of 0 to 100 and round the values to integers. In effect this converts a vector of size 1800 to a string of 400 integers. An example is shown in Figure \ref{fig:ir_input}.

\subsubsection{NMR Spectra}

\begin{figure}[h!]
    \centering
    \includegraphics[width=0.95\textwidth]{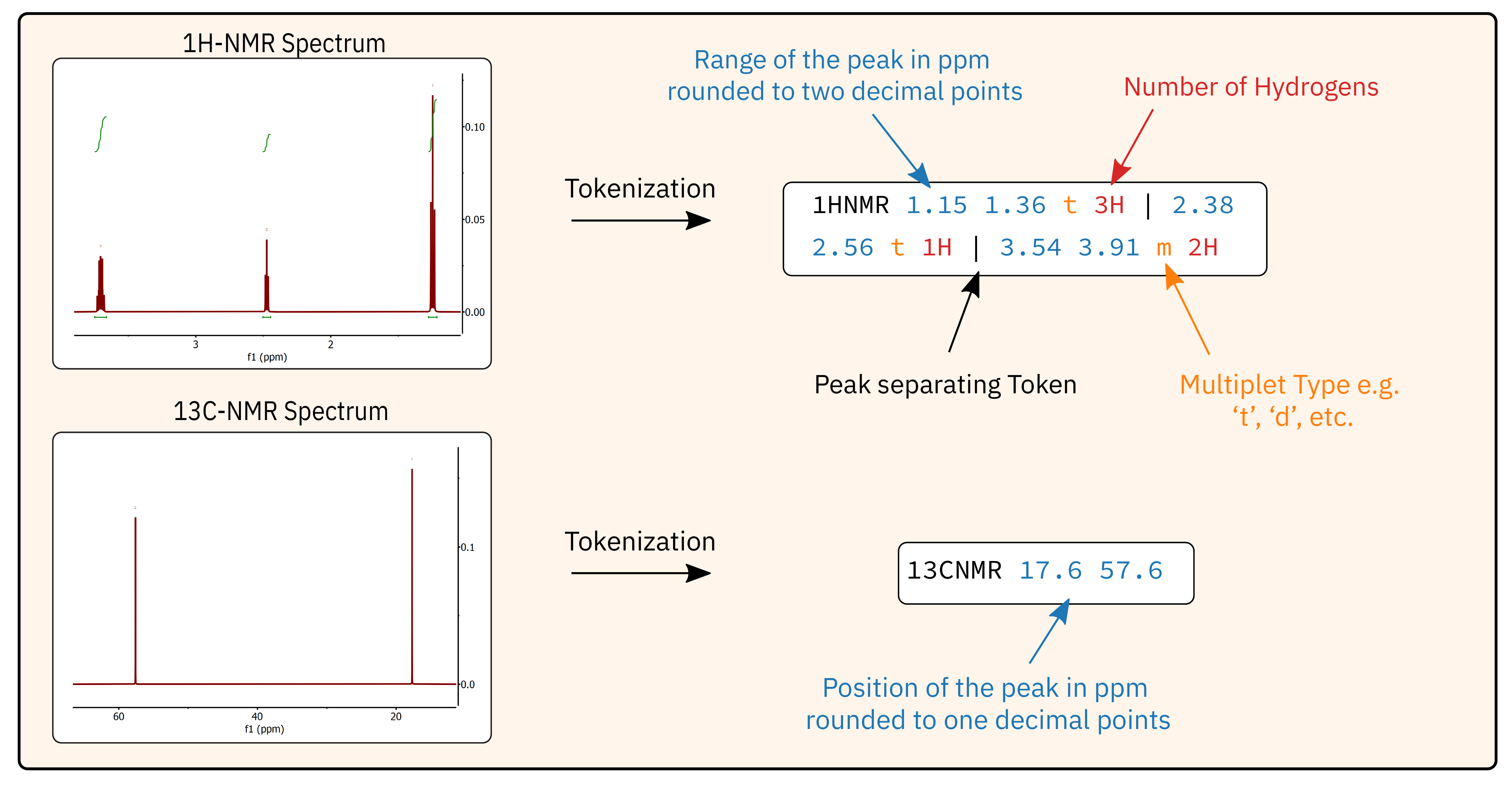}
    \caption{Summary of the tokenization process for NMR spectra. Top: Tokenization of an $^1$H-NMR spectrum following the Range representation. Bottom: Tokenization of a $^{13}$C-NMR spectrum}
    \label{fig:nmr_represenation}
\end{figure}

To tokenize $^1$H-NMR peaks, we proceed as follows: The position of the peak is rounded to the second decimal point, the type of multiplet (singlet, doublet, triplet, etc.) and the number of hydrogens are appended as second and third token respectively. All peaks are separated with a separating token (``\texttt{|}''). As an example a singlet at 1.239 ppm with an integral of 3 would become ``\texttt{1.24 s 3H |}'', with tokens separated by whitespaces. A string of the $^1$H NMR spectrum is built accordingly by concatenating the peaks starting with the lowest ppm and ending at the highest one. In addition, a prefix token is used to differentiate $^1$H- from $^{13}$C-NMR spectra. As an example an $^1$H-NMR with two peaks would be formatted as follows: ``\texttt{1HNMR 1.24 t 3H | 1.89 q 3H |}''.

$^{13}$C-NMR are formatted according to a simpler scheme. As the multiplet type and integration is not relevant for this type of spectrum the position of the peaks are rounded to one decimal point and tokenized accordingly. To illustrate this, a typical NMR spectrum is tokenized as follows: ``\texttt{13CNMR 12.1 27.8 63.5}''. 

Both methods are illustrated in Figure \ref{fig:nmr_represenation}. In case both the $^1$H- and $^{13}$C-NMR are used as input both are concatenated. 
\newpage
\subsubsection{MS/MS Spectra}

\begin{figure}[h!]
    \centering
    \includegraphics[width=0.95\textwidth]{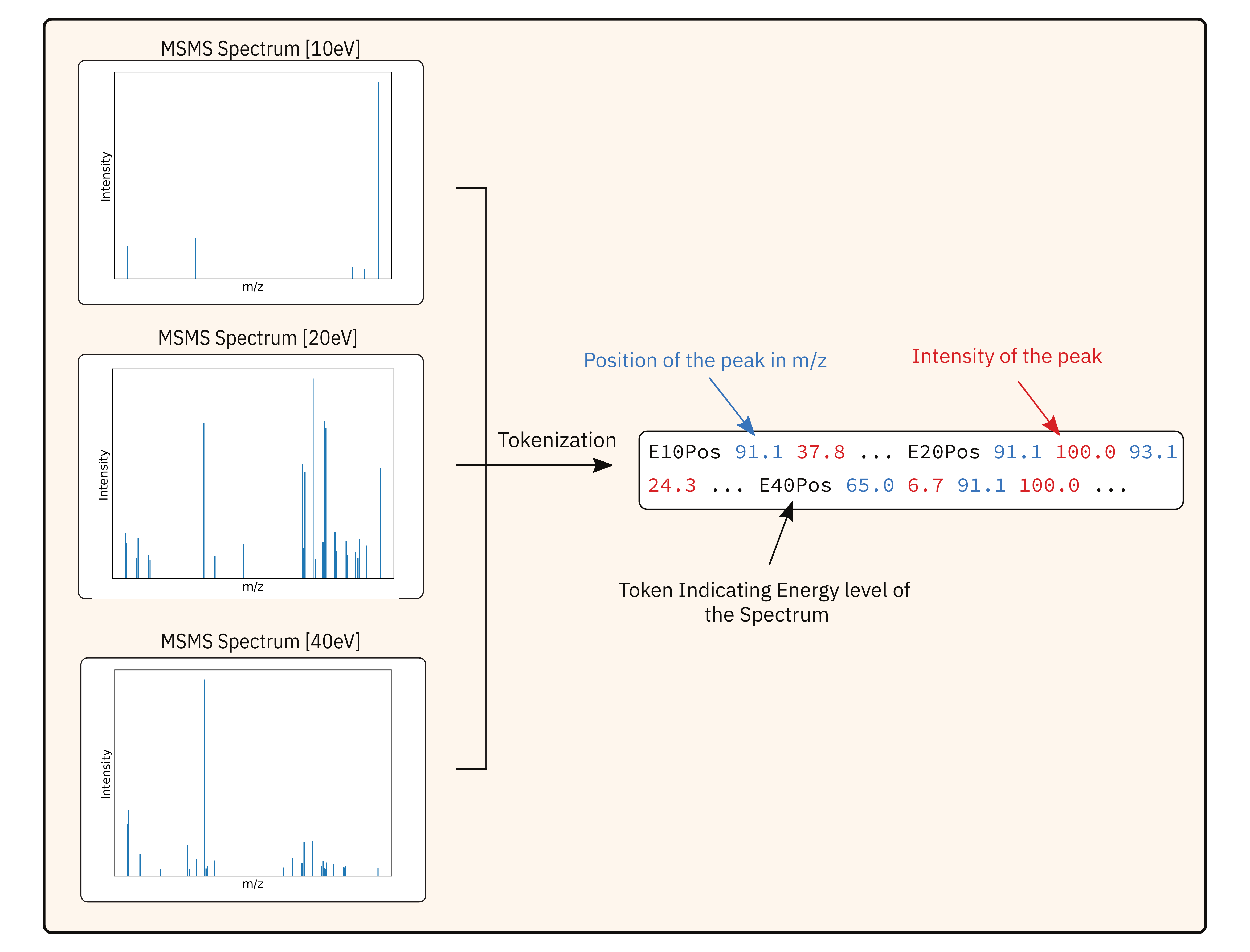}
    \caption{Tokenization of the MS/MS spectra: We include the position and intensity of each peak in the structured text represenation and concatenate the spectra resultant from the different ionisation energies.}
    \label{fig:msms_represenation}
\end{figure}

The generated MS/MS consist of stick spectra, i.e. a list of the position in m/z and the intensity of each peak. We convert this stick spectrum into a structured text representation by listing each peak followed by it's intensity. Both the m/z and intensity are rounded to one decimal point. We concatenate the MS/MS generated at the three different ionisation energies as shown in Figure \ref{fig:msms_represenation}.

\subsection{Zero Shot on van Bramer et. al.}
\label{app:zeroshot}
We evaluate the zero-shot performance of the models trained to predict the structure from the spectra on the experimental van Bramer dataset. The results are shown in Table \ref{tab:zeroshot} below:

\begin{table}[h!]
    \renewcommand{\arraystretch}{1.2}
    \caption{
        Zero shot accuracy (Top--1, Top--5, and Top--10) of transformer models trained on simulated data to predict the chemical structure from different spectra evaluated on the experimental van Bramer et. al. dataset.
    }
    \label{tab:zeroshot}
    \begin{center}
    \begin{tabular}{lrrr}
    \toprule
         & Top--1\% & Top--5\% & Top--10\% \\
        \midrule
        IR & 5.03 & 11.96 & 17.61 \\ 
        MS/MS (Negative, CFM-ID) & 0.0 & 0.0 & 0.0 \\
        MS/MS (Positive, CFM-ID) & 0.0 & 0.0 & 0.0 \\
        MS/MS (Positive, Scarf) & 8.81 & 25.79 & 36.47 \\
        MS/MS (Positive, Iceberg) & 5.66 & 11.94 & 13.83 \\
        $^{13}$C-NMR & 58.49 & 69.81 & 72.96 \\
        $^{1}$H-NMR & 42.77 & 55.35 & 59.75 \\ 
        $^{1}$H-NMR + $^{13}$C-NMR & 50.31 & 63.98 & 67.08 \\
        \bottomrule
    \end{tabular}
    \end{center}
\end{table}

\newpage
\subsection{Compute Resources used for simulating spectra}

In the following, we outline the compute resources used to generate the dataset. All simulations were run on AMD EPYC 7452 CPUs.
\begin{itemize}
    \item \textbf{IR}: 500 CPU cores with 1TB RAM for ca. 46 days
    \item \textbf{$^1$H-NMR:} 200 CPU cores with 400GB RAM for ca. 15 days.
    \item \textbf{$^{13}$C-NMR:} 200 CPU cores with 400GB RAM for ca. 14 days.
    \item \textbf{MSMS-CFM ID:} 80 CPU cores with 160GB RAM for ca. 7 days.
    \item \textbf{MSMS-Iceberg:} 16 CPU cores with 32GB RAM for ca. 3 days.
    \item \textbf{MSMS-Scarf:} 16 CPU with 32GB RAM cores for ca. 1.5 days 
\end{itemize}

\newpage

\section*{Checklist}

\begin{enumerate}

\item For all authors...
\begin{enumerate}
  \item Do the main claims made in the abstract and introduction accurately reflect the paper's contributions and scope? \answerYes{}
    
  \item Did you describe the limitations of your work?
  \answerYes{}
  \item Did you discuss any potential negative societal impacts of your work?
    \answerNA{}
  \item Have you read the ethics review guidelines and ensured that your paper conforms to them?
    \answerYes{}
\end{enumerate}

\item If you are including theoretical results...
\begin{enumerate}
    \item Did you state the full set of assumptions of all theoretical results?
    \answerNA{}
        \item Did you include complete proofs of all theoretical results?
    \answerNA{}
\end{enumerate}

\item If you ran experiments (e.g. for benchmarks)...
\begin{enumerate}
  \item Did you include the code, data, and instructions needed to reproduce the main experimental results (either in the supplemental material or as a URL)?
     \answerYes{}
  \item Did you specify all the training details (e.g., data splits, hyperparameters, how they were chosen)?
     \answerYes{}
	\item Did you report error bars (e.g., with respect to the random seed after running experiments multiple times)?
    \answerYes{}
	\item Did you include the total amount of compute and the type of resources used (e.g., type of GPUs, internal cluster, or cloud provider)?
    \answerYes{}
\end{enumerate}

\item If you are using existing assets (e.g., code, data, models) or curating/releasing new assets...
\begin{enumerate}
  \item If your work uses existing assets, did you cite the creators?
    \answerYes{}
  \item Did you mention the license of the assets?
    \answerYes{}
  \item Did you include any new assets either in the supplemental material or as a URL?
    \answerYes{}
  \item Did you discuss whether and how consent was obtained from people whose data you're using/curating?
     \answerNA{}: The USPTO dataset has "CC0" license
  \item Did you discuss whether the data you are using/curating contains personally identifiable information or offensive content?
    \answerNA{}
\end{enumerate}
\item If you used crowdsourcing or conducted research with human subjects...

\begin{enumerate}
\item  Did you include the full text of instructions given to participants and screenshots, if applicable?
 \answerNA{}
\item Did you describe any potential participant risks, with links to Institutional Review Board (IRB) approvals, if applicable?
\answerNA{}
\item  Did you include the estimated hourly wage paid to participants and the total amount spent on participant compensation? 
\answerNA{}
\end{enumerate}
\end{enumerate}

\end{document}